\documentclass[aps,pre,twocolumn,groupedaddress,showpacs,superscriptaddress,amssymb,amsmath]{revtex4-1}
\usepackage{graphicx}
\newcommand{\be}{\begin{equation}}
\newcommand{\ee}{\end{equation}}
\begin{document}


\title{Counting statistics of heat transport in harmonic junctions -- transient and steady states}



\author{Bijay Kumar Agarwalla} 
\affiliation{Department of Physics and Center for Computational Science and Engineering, National University of Singapore, Singapore, 117542, Republic of Singapore}
\author{Baowen Li}  
\affiliation{Department of Physics and Center for Computational Science and Engineering, National University of Singapore, Singapore, 117542, Republic of Singapore}
\affiliation{NUS Graduate School for Integrative Sciences and Engineering, Singapore 117456, Republic of Singapore}
\affiliation{Centre for Phononics and Thermal Energy Science, Department of Physics, Tongji University, 200092 Shanghai, China} 
\author{Jian-Sheng Wang} 
\affiliation{Department of Physics and Center for Computational Science and Engineering, National University of Singapore, Singapore, 117542, Republic of Singapore}


\date{21 November 2011}

\begin{abstract}
We study the statistics of heat transferred in a given time interval $t_M$, through a finite harmonic chain, called the center $(C)$, which is connected with two heat baths, the left $(L)$ and the right $(R)$, that are maintained at two different temperatures. The center atoms are driven by an external time-dependent force. We calculate the cumulant generating function (CGF) for the heat transferred out of the left lead, $Q_L$, based on two-time measurement concept and using nonequilibrium Green's function (NEGF) method. The CGF can be concisely expressed in terms of Green's functions of the center and an argument-shifted self-energy of the lead. The expression of CGF is valid in both transient and steady state regimes. We consider three different initial conditions for the density operator and show numerically, for one-dimensional (1D) linear chains, how transient behavior differs from each other but finally approaches the same steady state, independent of the initial distributions. We also derive the CGF for the joint probability distribution $P(Q_L,Q_R)$, and discuss the correlations between $Q_L$ and $Q_R$. We calculate the total entropy flow to the reservoirs. In the steady state we explicitly show that the CGF obeys steady state fluctuation theorem (SSFT).  Classical results are obtained
by taking $\hbar \to 0$. The method is also applied to the counting of the electron
number and electron energy, for which the associated self-energy is obtained from the usual lead self-energy by multiplying a phase or shifting the contour time, respectively. 
\end{abstract}


\maketitle


\section{Introduction}
Nonequilibrium systems are common in nature because they are, in general, subject to thermal gradients, chemical potential gradients or may be triggered by time-dependent forces. Heat transport is one such example of nonequilibrium systems where the heat carriers could be electrons, phonons, magnons, etc. To study heat transport in phononic systems, one considers a finite junction part, which can be an insulator, connected with two heat baths that are maintained at different temperatures. In the past decade, the main focus was on the calculation of the steady state heat current or heat flux flowing through the junction part from the leads 
\cite{Caroli,Meir,Rego-Kirczenow-1998,Segal,Mingo-Yang-2003,Yamamoto-2006,Dhar1,Wang-prb06,Wang-pre07,WangJS-europhysJb-2008}. For diffusive systems, the answer is given by Fourier's law \cite{Dhar, Lepri, Lebowitz} which is true only in the linear response regime, i.e., when the temperature difference between the baths is small. However for harmonic or ballistic systems, the heat current is given by a Landauer-like formula \cite{Rego-Kirczenow-1998,Dhar1,WangJS-europhysJb-2008} which was first derived for electronic transport. Landauer formula on the contrary to the Fourier's law is true for arbitrary temperature differences between the leads. No such explicit expression for current is known for transient states.  In recent times, several works \cite{eduardos-paper,eduardos-paper1} followed to answer what happens to current in the transient regime. This is an important question both from the theoretical and experimental points of view.  

Much attention has been given to phonon transport, in particular on thermal devices and on controlling heat flow \cite{baowen-review}. With the advent of technology it is now possible to study transport problems and observe a single mode of vibration in small systems with few degrees of freedom \cite{oconell}. These systems shows strong thermal fluctuations which play an important role because thermal fluctuations can lead to instantaneous heat transfer from colder to hotter lead. It is therefore necessary to talk about the statistical distribution of heat flux for these systems. In the electronic literature the distribution $P(Q_L)$ of the charge $Q_L$, flowing from the left lead to the junction part, was answered by calculating the corresponding CGF, ${\cal Z}(\xi)= \langle e^{i \xi Q_L} \rangle$, and is given by the celebrated Levitov-Lesovik formula \cite{Levitov,Levitov1,Levitov-PRB}. This methodology is also known as the {\it full counting statistics} \cite{otherworks,otherworks1,otherworks2,Klich,Pilgram,Bagrets,Gogolin,Urban,Gutman,fluct-theorems}
in the field of electronic transport. Experimentally the electron counting statistics has been measured in quantum-dot systems \cite{Flindt-etal-2009,Gustavsson-etal-2006}. However few experiments have been done for phonons \cite{measurement}. In the phononic case Saito and Dhar \cite{Saito-Dhar-2007} gave an explicit expression of the CGF.  Ren et al.\ gave a result for two-level systems \cite{ren-jie}. Full counting statistics of energy fluctuations in a driven quantum resonator is studied by Clerk \cite{AAClerk-2011}. The main focus in these papers was on the long-time limit and SSFT \cite{noneq-fluct,fluct-theorem-1,fcs-bijay,kundu,Esposito-review-2009, Hanggi2}. Using the NEGF method \cite{schwinger-keldysh,rammer86} and two-time measurement \cite{Esposito-review-2009, Hanggi2,Hanggi3,Hanggi4} concept, Wang~et al.\ \cite{fcs-bijay} gave an explicit expression for the CGF which is valid for both transient and steady state regimes.   

In this paper, we extend our previous work in Ref.~\onlinecite{fcs-bijay} and derive the CGF in a more general scenario, i.e., in the presence of both the temperature difference and the time-dependent driving force. We analyze the cumulants of heat $Q_L$ for three different initial conditions of the density operator and study the effects on both transient and steady state regimes. We also derive the CGF for the joint probability distribution of left and right lead heat $P(Q_L,Q_R)$ which help us to obtain the correlations between $Q_L$ and $Q_R$. By calculating CGF for $P(Q_L,Q_R)$ we can immediately obtain the CGF for the total entropy that flows to the leads. We present analytical expressions of the CGF's in the steady state and discuss the SSFT. Our method can be easily generalized for multiple heat baths.

The plan of the paper is as follows. We start in Sec.~II by introducing our model. Then in Sec.~III we define current and corresponding quantum heat operator followed by the definition of CGF for $Q_L$ using the two-time measurement concept, in Sec.~IV. In Sec.~V we derive the CGF using Feynman's path integral method and in Sec.~VI we use Feynman's diagrammatic technique to derive the CGF. We discuss the steady state result and fluctuation theorems in Sec.~VII. In Sec. VIII and IX we discuss how to calculate the CGF's numerically in transient regime and give numerical results for one-dimensional (1D) linear chain model, connected with Rubin heat baths, for three different initial conditions of the density operator. Then in Sec.~X we obtain the CGF for joint probability distribution of heat transferred $P(Q_L,Q_R)$ and discuss correlations and total entropy flow. In. Sec.~XI we give the long-time limit expression for the driven part of the full CGF. In Sec.~XII we discuss another definition of generating function due to Nazarov and discuss the corresponding long-time limit.  We found that using this definition, the generating function 
does not obey the Gallavotti-Cohen (GC) fluctuation symmetry. 
We conclude with a short discussion in Sec.~XIII. Few appendices give some details of technique nature. In particular, an electron system of a tight-binding model is treated using our method.

\section{The model}
Our model consists of a finite harmonic junction part, which we denote by $C$, coupled to two heat baths, the left ($L$) and the right ($R$), kept at two different temperatures $T_L$ and $T_R$, respectively. To model the heat baths, we consider an infinite collection of coupled harmonic oscillators. We take the three systems to be decoupled initially and to be described by the Hamiltonians,
\begin{equation}
{\cal H}_\alpha = \frac{1}{2} p_\alpha^T p_\alpha + \frac{1}{2} u_\alpha^T K^\alpha u_\alpha, \quad \alpha=L, C, R,
\end{equation}
for the left, right, and the finite central region. The leads are assumed to be semi-infinite. Masses are absorbed by defining $u = \sqrt{m}\,
x$. $u_\alpha$ and $p_\alpha$ are column vectors of coordinates and momenta. $K^\alpha$ is the spring constant matrix of region $\alpha$.
Couplings of the center region with the leads are turned on either adiabatically from time $t=-\infty$, or switched on abruptly at $t=0$.
The interaction Hamiltonian takes the form 
\begin{equation}
{\cal H}_{{\rm int}} = u_L^T \, V^{LC} \,u_{C} + u_R^T\, V^{RC}\, u_{C}.
\end{equation}
For $t>0$, an external time-dependent force is applied only to the center atoms, which is of the form
\begin{equation}
{\cal V}_C(t) =-f^T(t)\,u_C,
\end{equation}
where $f(t)$ is the time-dependent force vector. The driving force couples only with the position operators of the center. The force can be in the form of electromagnetic field. Coupling of this form helps us to obtain analytical solution for the CGF of heat flux.
So the full Hamiltonian for $t>0$ (in the Schr\"odinger picture) is
\begin{equation}
{\cal H}(t) = {\cal H}(0^{-}) + {\cal V}_C(t) = {\cal H}_C + {\cal H}_L+ {\cal H}_R+ {\cal H}_{\rm int}+ {\cal V}_C(t).
\end{equation}
In the next section we will define current operator and the corresponding heat operator based on this Hamiltonian.
 
\section{Definition of current and heat operators}
It is possible to define the current operator ${\cal I}$ depending on where we want to measure the current. Here we consider the current flowing from the left lead to the center system and ${\cal I}_{L}$ is defined (in Heisenberg picture) as
\begin{equation}
{\cal I}_{L}(t) = - \frac{d{\cal H}^{H}_L(t)}{dt}=\frac{i}{\hbar}[{\cal H}^{H}_L(t), {\cal H}_H(t)]= p_L^{T}(t)\,V^{LC}\,u_C(t),
\label{current}
\end{equation}
where ${\cal H}_H(t)$ is the (time-dependent) Hamiltonian in the Heisenberg picture at time $t$. The corresponding heat operator can be written down as
\begin{equation}
\label{eq-hatQ}
{\cal Q}_{L}(t)=\int_0^t {\cal I}_{L}(t')\,dt' = {\cal H}_L(0)-{\cal H}^{H}_L(t),
\end{equation}
where ${\cal H}_L \big[={\cal H}_L(0)\big]$ is the Schr\"odinger operator of the free left lead and 
\begin{equation}
{\cal H}^{H}_L(t)= {\cal U}(0,t)\, {\cal H}_L\, {\cal U}(t,0),
\end{equation}
and ${\cal U}(t,t')$ is the evolution operator corresponding to the full Hamiltonian ${\cal H}(t)$ and satisfies the Schr\"odinger equation
\begin{equation}
i\hbar \,{ \partial {\cal U}(t,t') \over \partial t} = {\cal H}(t)\, {\cal U}(t,t').
\end{equation}
The formal solution of this equation is (assuming $t \geq t'$)
\begin{equation}
{\cal U}(t,t')=T \exp\left\{ - \frac{i}{\hbar} \int_{t'}^t {\cal H}(\bar{t})\, d\bar{t} \right\},
\label{eq-unitary}
\end{equation}
where $T$ is the time-order operator where time increases from right to left. Also ${\cal U}^{\dagger}(t,t')={\cal U}(t',t)$. $Q$ of non-calligraphic font will be a classical variable. 

In the following section we derive the CGF based on this definition of heat operator and using two-time measurement scheme.

\section{Definition of the generating function for heat operator}
Our primary interest here is to calculate the moments or cumulants of the heat energy transferred in a given time interval $t_M$. Hence, it is advantageous to calculate the generating function instead of calculating moments directly. Since ${\cal Q}_{L}$ is a quantum operator, there are subtleties as to how exactly the generating function should be defined. Naively we may use $\langle e^{i \xi {\cal Q}_{L}} \rangle $. But this definition fails the fundamental requirement of positive definiteness of the probability distribution.

Here we will give two different definitions that are used to calculate the generating function for such problem. The first definition comes from the idea of two-time measurements and based on this concept the CGF can be written down as
\begin{equation}
{\cal Z}(\xi)=\langle e^{i \xi {\cal H}_L} \, e^{-i \xi {\cal H}^{H}_L(t) } \rangle'
\label{eq-Z-two-time}
\end{equation}
which we will discuss in great detail in this section.

The second definition of the CGF is
\be
{\cal Z}_1(\xi) = \langle \bar{T} e^{i\xi {\cal Q}_{L}/2} T e^{i\xi {\cal Q}_{L}/2} \rangle,
\label{eq-Z1-Nazarov}
\ee
where $\bar{T}$ is the anti-time order operator. The time (or anti-time) order is meant to apply to the integrand when the
exponential is expanded and ${\cal Q}_{L}$ is expressed as integral over ${\cal I}_{L}$ as in Eq.~(\ref{eq-hatQ}). This definition is used by Nazarov et al.\ \cite{otherworks,otherworks1} mostly for the electronic transport case. In the last section we will show how this generating function can be derived starting from ${\cal Z}(\xi)$ under a particular approximation and will also give explicit expression for ${\cal Z}_{1}(\xi)$ in the long-time limit. 

In the following we will discuss the idea of two-time measurement and derive the corresponding CGF ${\cal Z}(\xi)$.
 
\subsection{Two-time measurement}
The heat operator in Eq.~(\ref{eq-hatQ}) depends on the left-lead Hamiltonian ${\cal H}_{L}$ at time 0 and $t$. The concept of two-time measurement implies the measurement of a certain operator (in this case ${\cal H}_{L})$ at two different times. Here the measurement is in the sense of quantum measurement
of von Neumann \cite{neumann}.   

Let us first assume that the full system is in a pure state $|\Psi_0\rangle$ at $t=0$.
We want to do measurement of the energy associated with the operator
${\cal H}_L$.  According to quantum mechanics, the result of a measurement can
only be an eigenvalue of the (Schr\"odinger) operator ${\cal H}_L$ and the wave function
collapses into an eigenstate of ${\cal H}_L$.  Let
\begin{equation}
{\cal H}_L | \phi_a \rangle = a | \phi_a \rangle,\quad \Pi_a = |\phi_a\rangle \langle \phi_a  |,
\end{equation}
where $\Pi_a$ is the projector into the state $|\phi_a\rangle$ satisfying $\Pi_a^2 = \Pi_a$,
and $\sum_a \Pi_a = 1$.  We assume the eigenvalues are discrete (this is always so
if the lattice system is finite). After the measurement at time $t=0$, the wave function
is proportional to $\Pi_a | \Psi_0 \rangle$ if the result of the measurement is the energy $a$ and the probability of such event happen is $\langle \Psi_0 | \Pi_a^2 | \Psi_0 \rangle$. Let's propagate this state to time $t$ and do a second measurement of the lead
energy, finding that the result is $b$.  The wave function now becomes proportional to $\Pi_b \, {\cal U}(t,0) \, \Pi_a \,  | \Psi_0 \rangle$.
The joint probability of getting $a$ at time $0$ and $b$ at time $t$ is
the norm (inner product) of the above (unnormalized) state. 

If the initial state is in a mixed state, we add up the initial probability classically, i.e., if 
\begin{equation}
\rho(0) = \sum_k w_k | \Psi_0^k \rangle\langle \Psi_0^k |,\quad w_k > 0, \quad \sum_k w_k = 1,
\end{equation}
the joint probability distribution of two-time measurement output is
\begin{eqnarray}
P(b,a) &=& \sum_k w_k \langle \Psi_0^k | \,\Pi_a\, {\cal U}(0,t) \,
\Pi_b \,{\cal U}(t,0) \,\Pi_a \,| \Psi_0^k \rangle  \nonumber \\
&=& {\rm Tr} \bigl[ \Pi_a \,\rho(0) \, \Pi_a \, {\cal U}(0,t) \, \Pi_b \, {\cal U}(t,0)\bigr].
\end{eqnarray}
By definition, we see that $P(b,a)$ is a proper probability in the sense that $P(b,a) \ge 0$ and $\sum_{a,b} P(b,a) = 1$.
Then the generating function for $Q_L=a-b$ is defined as
\begin{eqnarray}
{\cal Z}(\xi) &=& \langle e^{i\xi (a-b)} \rangle =  \sum_{a,b} e^{i \xi(a-b)} P(b,a) \nonumber \\
&=&  \sum_{a,b} e^{i \xi (a-b)} {\rm Tr} \bigl[\Pi_a \, \rho(0) \, \Pi_a \, {\cal U}(0,t) \, \Pi_b \, {\cal U}(t,0) \bigr] \nonumber \\
& = & \langle e^{i \xi {\cal H}_L} \, e^{-i \xi {\cal H}^{H}_L(t) } \rangle' \nonumber \\
\label{eqZ2-def}
& = &  \langle e^{i \xi {\cal H}_L/2} \, e^{-i \xi {\cal H}^{H}_L(t) } \, e^{i \xi {\cal H}_L/2 } \rangle'.
\end{eqnarray}
where we define \cite{neumann}
\begin{equation}
\rho'(0) = \sum_{a} \Pi_a\, \rho(0) \Pi_a.
\label{projected}
\end{equation}
which we call as the {\it projected} density matrix.

If the initial state at $t=0$ is a product state  i.e., $\rho(0)=\rho(-\infty)=\rho_L \otimes \rho_C \otimes \rho_R $, where the left, center and right density
matrices are in equilibrium distributions corresponding to the 
respective temperatures:
$\rho_\alpha={e^{-\beta_\alpha {\cal{H}}_\alpha}}/{{\rm Tr} [e^{-\beta_\alpha
    {\cal H}_\alpha}]}$ for $\alpha=L,C,R$ and $\beta_{\alpha}=1/(k_{\rm B}
T_{\alpha})$, then the projection operators $\Pi_a$ do not play any role and $\langle....\rangle'={\rm Tr}\Bigl[\rho(-\infty)\cdots \Bigr]=\langle....\rangle$. 

Here we will derive the CGF for three different initial conditions:
\begin{itemize}
\item{Product initial state, i.e., $\rho(-\infty)$, which corresponds to sudden switch-on of the coupling between the leads and the center.}
\item{steady state as the initial state, i.e., $\rho(0)$, which we can obtain, starting with the decoupled Hamiltonians at $t=-\infty$, switch on the couplings between the center region and the leads, adiabatically upto time $t=0$.}
\item{{\it projected} density matrix $\rho'(0)$ considering $\rho(0)$ as the steady state, i.e., taking the effects of measurements into account.}
\end{itemize}   

In the following sections we will analytically show that the CGF's corresponding to different initial conditions reach the same steady state in the long-time limit and hence is independent of initial distributions. However for short time transient behavior depends significantly on initial conditions and also the measurements do play an important role. 

\section{calculation for ${\cal Z}(\xi)$ for initial states $\rho(0)$ and $\rho'(0)$}
In this section we will give detail derivation for ${\cal Z}(\xi)$, using Feynman path-integral formalism, for two different initial density operators $\rho(0)$ and $\rho'(0)$. 

\subsection{Removing the projection $\Pi_a$ at $t=0$}
The projection by $\Pi_a$ at $t=0$ Eq.~(\ref{projected}) to the density matrix creates a problem
for formulation in path integrals.  We can remove it following Ref.~\onlinecite{Esposito-review-2009} by putting it into part of an evolution of ${\cal H}_L$, just like the factor associated with the generating function variable $\xi$, with a price we have to pay, introducing another integration variable $\lambda$. The key observation is that we can represent the projector by the Dirac
$\delta$ function
\begin{eqnarray}
\Pi_a & \propto & \delta(a-{\cal H}_L)  \nonumber \\
&=& \int_{-\infty}^\infty \frac{d\lambda}{2\pi} \,e^{-i\lambda(a-{\cal H}_L)}.
\label{eq-delta}
\end{eqnarray}
For this to make sense, we assume the spectrum of the energy of
${\cal H}_L$ is continuous, which is valid if we take the large size limit first.
Identifying $\Pi_a$ as $\delta(a-{\cal H}_L)$ with a continuous variable
$a$ introduces an constant proportional to the Dirac $\delta(0)$ to $\rho'(0)$, since
$\Pi_a$ is now normalized as $\Pi_a \Pi_b = \delta(a-b) \Pi_a$.
However, this constant can be easily fixed by the condition
${\cal Z}(0)=1$.  So using $\Pi_a = \delta(a -{\cal H}_L)$ will not cause
difficulty.  

Substituting the Fourier integral representation into $\rho'$ we obtain
\begin{eqnarray}
\rho'(0) &\propto&\int da\, \Pi_a\, \rho(0) \Pi_a  \\
&=& \int \frac{d\lambda}{2\pi} e^{i\lambda {\cal H}_L} \rho(0) e^{-i\lambda {\cal H}_L}.
\end{eqnarray}
Using the symmetric form of ${\cal Z}$, Eq.~(\ref{eqZ2-def}), we have
\begin{eqnarray}
{\cal Z}(\xi)&\propto& \int \frac{d\lambda}{2\pi} {\rm Tr} \bigl\{ 
\rho(0) \,{\cal U}_{\xi/2-\lambda}(0,t)\, {\cal U}_{-\xi/2-\lambda}(t,0) \bigr\}  \nonumber \\
&=&  \int \frac{d\lambda}{2\pi}\; {\cal Z}(\xi, \lambda),
\label{eq-Uxilambda}
\end{eqnarray}
where ${\cal U}_{x}(t,t')$ is the modified evolution operator of an effective Hamiltonian given by
\begin{equation}
{\cal H}_{x}(t) = e^{i x {\cal H}_L} {\cal H}(t) e^{-i x {\cal H}_L},
\end{equation}
where $x$ is a real parameter which in this case is $\xi/2 -\lambda$ and $-\xi/2 -\lambda$. Finally ${\cal U}_{x}(t,t')$ is given by ($t \geq t'$)
\begin{eqnarray}
{\cal U}_{x}(t,t') &=& e^{i x {\cal H}_L} {\cal U}(t,t') e^{-i x {\cal H}_L} \nonumber \\
&=& \sum_{n=0}^\infty \left(-\frac{i}{\hbar} \right)^n \int_{t'}^t dt_1 \int_{t'}^{t_1} dt_2 \cdots \int_{t'}^{t_{n-1}}dt_n
\nonumber \\ 
&& \times e^{i x {\cal H}_L} {\cal H}(t_1) {\cal H}(t_2) \cdots {\cal H}(t_n) e^{-i x {\cal H}_L} \nonumber \\
\label{eq-Ulambda}
&=& T \exp\left\{ - \frac{i}{\hbar} \int_{t'}^t {\cal H}_x(t') dt' \right\}.
\end{eqnarray}
It is important to note that substituting $\lambda=0$ in ${\cal Z}(\xi,\lambda)$ gives us the initial density matrix $\rho(0)$. 

Now we will give an explicit expression of the modified Hamiltonian ${\cal H}_x$ which helps us to calculate the CGF using path integral.

\subsection{The expression for ${\cal H}_{x}$}
The modified Hamiltonian is the central quantity for calculating CGF. It is the Heisenberg evolution of the full Hamiltonian ${\cal H}(t)$ (in Schr\"odinger picture) with respect to ${\cal H}_L$. Since ${\cal H}_L$ commutes with
every term $\tilde{\cal H}$ where ${\cal H}(t)= {\tilde {\cal H}} + u_L^T V^{LC} u_C$, except the coupling term $u_L^T V^{LC} u_C$, we can write
\begin{eqnarray}
{\cal H}_{x}(t) &=& e^{i x {\cal H}_L} {\cal H}(t) e^{-i x {\cal H}_L} \nonumber \\
& = &  e^{i x {\cal H}_L} \bigl({\tilde {\cal H}} + u_L^T V^{LC} u_C \bigr) e^{-i x {\cal H}_L} \nonumber \\
& = & {\cal H}(t) + \bigl( u_L(\hbar x) - u_L\bigr)^T V^{LC} u_C,
\end{eqnarray}
where $u_L(\hbar x) = e^{i x {\cal H}_L} u_L e^{-i x {\cal H}_L}$ is the free left lead
Heisenberg evolution to time $t = \hbar x$.  $u_L(\hbar x)$ can be obtained
explicitly as
\begin{equation}
u_L(\hbar x) = \cos(\sqrt{K_L} \hbar x) u_L + \frac{1}{\sqrt{K_L}} \sin( \sqrt{K_L} \hbar x) p_L.
\end{equation} 
The matrix $\sqrt{K_L}$ is well-defined as the matrix $K_L$ is positive definite. $u_L$ and $p_L$ are the initial conditions at $t=0$.
The final expression for ${\cal H}_x(t)$ is 
\begin{equation}
{\cal H}_x(t) = {\cal H}(t) + \bigl[ u_L^T {\cal C}(x) + p_L^T {\cal S}(x) \bigr]u_C,
\label{modified}
\end{equation}
where 
\begin{eqnarray}
\label{eq-C}
{\cal C}(x) &=& \bigl(\cos(\hbar x \sqrt{K_L}) -I\bigr) V^{LC}, \\
\label{eq-S}
{\cal S}(x) &=& (1/\sqrt{K_L}) \sin( \hbar x \sqrt{K_L}) V^{LC}.
\end{eqnarray}
The effective Hamiltonian now has two additional term with respect to the full ${\cal H}(t)$. The term $u_L^T {\cal C}(x)u_C $ is like the harmonic coupling term which modifies the coupling matrix $V^{LC}$.

In the following we calculate the two parameter generating function ${\cal Z}(\xi,\lambda)$ using Eq.~(\ref{eq-Uxilambda}).

\subsection{Expression for ${\cal Z}(\xi,\lambda)$}
\begin{figure}[t]
\includegraphics[width=0.5\columnwidth]{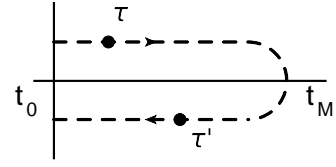}%
\caption{\label{fig1}The complex-time contour in the Keldysh formalism. The path of the contour begins at time $t_{0}$, goes to time $t_M$, and then goes back to time $t=t_{0}$. $\tau$ and $\tau'$ are complex-time variables along the contour. $t_{0}=-\infty$ and $0$ corresponds to Keldysh contour K and C, respectively.} 
\end{figure}

The expression for ${\cal Z}(\xi,\lambda)$ can be written down on the contour as (see Fig.~\ref{fig1})
\begin{equation}
{\cal Z}(\xi,\lambda) = {\rm Tr} \Bigl[ \rho(0) T_c e^{-\frac{i}{\hbar} \int_C {\cal H}^{x}(\tau) d\tau} \Bigr].
\label{eq-Z-contour} 
\end{equation} 
where $T_c$ is the contour-ordered operator which orders operators according to their contour time argument, earlier contour time places an operator to the right. The contour function $x(\tau)$ is defined as 0 whenever $t < 0$ or $t>t_M$, and when $0 < t < t_M$, i.e., within the measurement time interval, for upper branch of the contour $x^{+}(t) = -\xi/2 - \lambda$, and for lower branch $x^{-}(t) = \xi/2 - \lambda$. 
 
For the moment, let us forget about the other lead and concentrate only on the left lead and center.
The effect of other lead simply modifies the self-energy of the
leads additively, according to Feynman and Vernon \cite{Feynman-Vernon-1963}. Using Feynman path integral technique we can write
\begin{equation}
{\cal Z}(\xi,\lambda)= \int {\cal D}[u_C] {\cal D}[u_L] \rho(-\infty) e^{(i/\hbar) 
\int_K d\tau ( {\cal L}_C + {\cal L}_L + {\cal L}_{LC} ) }.
\label{eq-Z-keldysh}
\end{equation}
Note that in Eq.~(\ref{eq-Z-contour}), the 
contour $C$ is from 0 to $t_M$ and back, while that in Eq.~(\ref{eq-Z-keldysh}) is on the
Keldysh contour $K$, that is, from $-\infty$ to $t_M$ and back to take into account
of adiabatic switch on, replacing $\rho(0)$ by $\rho(-\infty)$.   Their relation
is
\begin{equation}
\rho(0) = {\cal U}(0, -\infty) \rho(-\infty) {\cal U}(-\infty, 0).
\end{equation}
We can identify the Lagrangian's as 
\begin{eqnarray}
{\cal L} &=& {\cal L}_L + {\cal L}_C + {\cal L}_{LC}, \nonumber \\
{\cal L}_L &=& \frac{1}{2}\dot{u}_L^2 - \frac{1}{2} u_L^T K^L u_L, \nonumber \\
{\cal L}_C &=& \frac{1}{2} \dot{u}_C^{2}  +f^T u_C -\, \frac{1}{2} u_C^T \bigl(K^C-{\cal S}^T{\cal S}\bigr) u_C, \nonumber \\
{\cal L}_{LC} &=& - \dot{u}^T_L {\cal S} u_C - u_L^T \bigl(V^{LC} + {\cal C}\bigr) u_C.
\label{Lagrangian}
\end{eqnarray}
For notational simplicity, we have dropped the argument $\tau$.
The vector or matrices $f$, ${\cal C}$, and ${\cal S}$ are parametrically dependent on
the contour time $\tau$.  They are zero except on the interval
$0 < t < t_M$.  $f$ is the same on the upper and lower branches,
while ${\cal C}$ and ${\cal S}$ take different values depending on $x(\tau)$.

Now the lead part can be integrated out by performing Gaussian integral \cite{Feynman-Vernon-1963}. Since the
coupling between the lead and the center is linear, it is plausible that the result will be
a quadratic form in the exponential, i.e., another Gaussian.  To find exactly
what it is, we convert the path integral back to the
interaction picture (with respect to ${\cal H}_L$) operator form and evaluate the expression by standard perturbative expansion. The only difference is that now the coupling with the center involving both $u_L$ and $\dot{u}_L$. The result for the influence
functional is given by \cite{Stockburger}
\begin{eqnarray}
I_L[u^C(\tau)] &\equiv & \int {\cal D}[u_L] \rho_L(-\infty) 
e^{\frac{i}{\hbar} \int d\tau ({\cal L}_L + {\cal L}_{LC}) } \nonumber  \\
&=& {\rm Tr}\Bigl[\frac{e^{-\beta_L H_L}}{Z_L} T_c e^{ -\frac{i}{\hbar} \int d\tau {\cal V}_I(\tau) } \Bigr] \nonumber \\
&=& e^{-\frac{i}{2\hbar} \int d\tau \int d\tau' u_C^T(\tau) \Pi(\tau, \tau') u_C(\tau') }.\
\end{eqnarray}

In the influence functional, the contour function $u_C(\tau)$ is not a 
dynamical variable but a parametric function. ${\cal V}_I(\tau)$ is the interaction picture operator with respect to the Hamiltonian ${\cal H}_L$ and is given by
\begin{eqnarray}
{\cal V}_I(\tau) &=& p_L^T {\cal S} u_C + u_L^T(V^{LC} + {\cal C}) u_C+\frac{1}{2} u_C^T {\cal S}^T {\cal S} u_C \nonumber \\
\label{eq-SSterm} 
&=&u_L^T\bigl(\tau + \hbar x(\tau)\bigr)V^{LC}u_C + \frac{1}{2} u_C^T {\cal S}^T {\cal S} u_C.
\end{eqnarray}  

The important influence functional self-energy on contour is given by
\begin{eqnarray}
\Pi(\tau,\tau')= \Sigma_L^A(\tau,\tau') &+& \Sigma_L(\tau,\tau') + {\cal S}^T{\cal S}\,\delta(\tau, \tau'), \\
\Sigma_L^A(\tau,\tau') + \Sigma_L(\tau,\tau') &= & V^{CL} g_L\bigl(\tau + \hbar x(\tau), \tau'+\hbar x(\tau') \bigr) V^{LC} \nonumber \\
\label{eq-SAL}
&=&  \Sigma_L\bigl(\tau + \hbar x(\tau), \tau'+\hbar x(\tau') \bigr),
\label{eq-shifted-self-energy}
\end{eqnarray}
where we obtain a shifted self-energy $\Sigma_L\big(\tau+\hbar x(\tau),\tau'+\hbar x(\tau')\big)$ which is the usual self-energy of the lead in contour time with arguments
shifted by $\hbar x(\tau)$ and $\hbar x(\tau')$. We define the self-energy $\Sigma^A_L$ as the difference between the shifted self-energy and the usual one $\Sigma_L(\tau,\tau')$. $\Sigma^{A}_L$ turns out to be a central quantity for this problem as we will show that, the CGF ${\cal Z}$ can be concisely expressed in terms of the center Green's function $G_{0}$ and $\Sigma^{A}_L$.

Substituting the explicit expression for the influence functionals of both the left and right leads to the path integral expression given in Eq.~(\ref{eq-Z-keldysh}), we have
\begin{eqnarray}
{\cal Z}(\xi,\lambda)&=& \int {\cal D}[u_C] \rho_C(-\infty) e^{(i/\hbar) 
\int d\tau  {\cal L}_C } I_L[u_C] I_R[u_C]\nonumber \\
&=& \int {\cal D}[u_C] \rho_C(-\infty) e^{\frac{i}{\hbar}S_{\rm eff}},
\end{eqnarray}
where the effective action is given by
\begin{eqnarray}
S_{\rm eff} &=& \int d\tau \Bigl[ 
\frac{1}{2} \dot{u}_C^2 - \frac{1}{2} u_C^T K^C u_C 
+ f^T u_C \Bigr]  \\
& \!\!-\!& \frac{1}{2}\! \int d\tau \!\int d\tau'\! u_C^T(\tau) \bigl(
\Sigma(\tau, \tau') + \Sigma_L^A(\tau, \tau') \bigr) u_C(\tau'), \nonumber
\end{eqnarray}
where $\Sigma = \Sigma_L +\Sigma_R$, taking into account the effect of both the leads.  The ${\cal S}^T{\cal S}$ term in $I_{L}[u_C]$ cancels exactly with
the one in ${\cal L}_C$. We can perform an integration by part on the $\dot{u}^2$ term, assuming that the surface term
does not matter (since it is at $t=-\infty$), we can write the expression in a standard quadratic form
\begin{eqnarray}
S_{\rm eff} &=& \frac{1}{2} \int d\tau \int d\tau' u_C^T(\tau) D(\tau, \tau') u_C(\tau') \nonumber \\
&& \qquad  + \int f^T(\tau) u_C(\tau) d\tau. 
\end{eqnarray}
$D(\tau,\tau')$ is the differential operator and is given by
\begin{eqnarray}
D(\tau, \tau') &=& -I \frac{\partial^2}{\partial \tau^2} \delta(\tau, \tau')
- K^C  \delta(\tau, \tau') \nonumber \\
&& -\Sigma(\tau, \tau') - \Sigma_L^A(\tau, \tau') \nonumber \\
&=& D_0(\tau,\tau') - \Sigma^A_L(\tau,\tau').
\end{eqnarray}
The above equation defines the Dyson equation on Keldysh contour. The generating function is obtained by doing another Gaussian integration and is of the following form
\begin{equation}
{\cal Z} \propto {\rm det}(D)^{-1/2} e^{-\frac{i}{2\hbar} f^T D^{-1} f}.
\label{eq-Z-det}
\end{equation}
(The meaning of the determinant will be explained in Appendix~C). 
We define the Green's function $G$ and $G_{0}$ by $DG = 1$, and $D_0 G_0 = 1$, or more precisely
\begin{equation}
\int D(\tau, \tau'') G(\tau'', \tau') d\tau'' = I \delta(\tau, \tau'),
\end{equation}
and similarly for $G_0$. $G$ can be written in terms of $G_{0}$ in the following Dyson equation form
\begin{eqnarray}
\label{eq-Dyson-full}
G(\tau, \tau') &=& G_0(\tau,\tau') \\
&&\> + \int \int d\tau_1d \tau_2 
G_0(\tau, \tau_1) \Sigma_L^A(\tau_1,  \tau_2) G(\tau_2, \tau'). \nonumber
\end{eqnarray}

We view the differential operator
(integral operator) $D$ and $D^{-1}$ as matrices that are
indexed by space $j$ and contour time $\tau$. $f$ is a column vector.  The
exponential factor term can also be written as a trace, 
$f^T D^{-1} f = {\rm Tr}_{(j,\tau)} ( G f f^T)$.
We can fix the proportionality constant by noting that
${\cal Z}(\xi=0, \lambda=0) =1$.  Since when $\xi=0$, $\lambda=0$, we have
$x=0$ and thus $\Sigma_L^A(\tau,\tau') = \Sigma_L(\tau+x, \tau'+x') - \Sigma_L(\tau, \tau') = 0$, so $D=D_0$.  The properly normalized
CGF is 
\begin{equation}
\label{eq-Zxilam}
{\cal Z}(\xi, \lambda) = {\rm det}\bigl( D_0^{-1}D)^{-1/2}e^{-\frac{i}{2\hbar} f^T D^{-1} f}.
\end{equation}
We don't need to do anything for the exponential factor because of the following reason.
We note
\begin{eqnarray}
f^T G_0 f &=& \int \int d\tau d\tau' f(\tau)^T G_0(\tau, \tau') f(\tau') \\
&=& \sum_{\sigma,\sigma'} \int \int \sigma dt\, \sigma' dt' f(t)^T G_0^{\sigma\sigma'}(t,t') f(t').\nonumber
\end{eqnarray}
Since the driven force $f$ does not depend on the branch indices, i.e., 
$f^{+}(t) = f^{-}(t)$, we can take the summation inside and obtain
\begin{equation}
\sum_{\sigma\sigma'} \sigma\sigma' G^{\sigma\sigma'} = G_0^t + G_0^{\bar t} - G_0^> - G_0^< = 0.
\end{equation}

Finally making use of the formulas for operators or matrices 
${\rm det}(M) = e^{{\rm Tr} \ln M}$, and 
$\ln (1 - y) = -\sum_{k=1}^\infty \frac{y^k}{k}$ we can write the CGF in terms of $\Sigma^{A}_L$ for the {\it projected} initial condition case as,
\begin{eqnarray}
\ln {\cal Z}(\xi)&=& \lim_{\lambda \to \infty}\ln {\cal Z}(\xi,\lambda) \nonumber \\
 &=& \lim_{\lambda \to \infty} \left\{-\frac{1}{2} {\rm Tr}_{j,\tau} \ln ( 1 - G_0 \Sigma_L^A) - \frac{i}{2\hbar} {\rm Tr}_{j,\tau}( G f f^T) \right\} \nonumber \\
&=& \lim_{\lambda \to \infty} \sum_{n=1}^\infty \frac{1}{2n}
{\rm Tr}_{(j,\tau)} \Bigl[(G_0 \Sigma_L^A)^n \Bigr] \nonumber  - \frac{i}{2\hbar} f^T G f \nonumber \\
&=& \frac{1}{2} {\rm Tr}_{(j,\tau)}(G_0 \Sigma_L^A) + \frac{1}{4} {\rm Tr}_{(j,\tau)}(G_0 \Sigma_L^A G_0 \Sigma_L^A) + \cdots \nonumber \\
&& -\frac{i}{2\hbar} f^T G_0 \Sigma^A G_0 f + \cdots.
\label{projected_state}
\end{eqnarray}
This expression for CGF is valid for any transient time $t_M$ present in the self-energy
$\Sigma^A_{L}$ and is the starting point for the calculation in transient regime.  The notation ${\rm Tr}_{(j,\tau)}$ means trace both in
space index $j$ and contour time $\tau$ (see Appendix~C). In order to obtain ${\cal Z}(\xi)$ from ${\cal Z}(\xi,\lambda)$ we have to take the limit $\lambda \rightarrow \infty$ because ${\cal Z}(\xi, \lambda)$ approaches a constant as $| \lambda| \to \infty$ and hence
the value of the integral is dominated by the value at infinity. Since $\Sigma^{A}_{L}(\tau,\tau')=0$ for $\xi=0$ we have the correct normalization ${\cal Z}(0)=1$.

Similarly, for the steady state initial condition $\rho(0)$ the CGF is given by 
\begin{equation}
\ln {\cal Z}(\xi)=\lim_{\lambda \rightarrow 0} \ln {\cal Z}(\xi,\lambda)
\end{equation}
The difference in this two cases is in the matrix $\Sigma^{A}_L$. 

Similar relations also exist if we want to calculate the CGF for right lead heat operator ${\cal Q}_R$. In this case one has to do two-time measurement on the right lead corresponding to the Hamiltonian ${\cal H}_R$. The final formula for the CGF remains the same except $\Sigma^{A}_L$ should be replaced by $\Sigma^{A}_R$.
 
Now in order to calculate the cumulants $\langle \langle Q_{\alpha}^{n} \rangle \rangle$ with $\alpha=L,R$  we need to go to the real time using Langreth's rule \cite{rammer86}. In this case, it is more convenient to work with a Keldysh rotation (see Appendix~C) for the contour ordered functions while keeping
${\rm Tr}(ABC \cdots D)$ invariant.  The effect of the Keldysh rotation is to change any given matrix ${\cal D}^{\sigma\sigma'}(t,t')$,
with $\sigma, \sigma'=\pm$ for branch indices, to,
\begin{eqnarray}
\label{keldysh-rotation}
\breve{{\cal D}} &=& 
\left( \begin{array}{cc}
                             {\cal D}^r & {\cal D}^K \\
                             {\cal D}^{\bar K} & {\cal D}^a 
\end{array} \right)  \\
 &=&
\frac{1}{2} \left( \begin{array}{cc}
              {\cal D}^t - {\cal D}^< + {\cal D}^> - {\cal D}^{\bar t}, & {\cal D}^t + {\cal D}^{\bar t} + {\cal D}^< + {\cal D}^> \\
              {\cal D}^t + {\cal D}^{\bar t} - {\cal D}^< - {\cal D}^>, & {\cal D}^< - {\cal D}^{\bar t} + {\cal D}^{t} - {\cal D}^{>}  
                         \end{array} \right). \nonumber
\end{eqnarray}
In this case we define the quantities 
${\cal D}^r$, ${\cal D}^a$, ${\cal D}^K$, and ${\cal D}^{\bar{K}}$ as above.  In particular,
${\cal D}^{K} \neq {\cal D}^< + {\cal D}^>$, as one usually might thought it is. 

Using the above definition for the center Green's function $G_0$ we get 
\begin{equation}
\breve{G_0} = 
\left( \begin{array}{cc}
                             G_0^r & G_0^K \\
                             0 & G_0^a 
\end{array} \right). 
\end{equation}
The $G_0^{\bar K}$ component is 0 due to the standard relation
among Green's functions.  But the $\bar K$ components are not zero for
$\Sigma_L^A$ and $G_0$, as we will compute later.  

It is useful computationally to work in Fourier space even if 
there is no time translational invariance.  We define the two-frequency
Fourier transform by
\begin{equation}
\breve{A}[\omega, \omega'] =
\int_{-\infty}^{+\infty}\!\!\! dt \int_{-\infty}^{+\infty}\!\!\! dt' 
\breve{A}(t,t') e^{i(\omega t + \omega' t')}.
\label{two-time-FT}
\end{equation}
Since $\breve{G}_0$ is time-translationally invariant then,
\begin{equation}
\breve{G}_0[\omega, \omega'] = 2\pi \delta(\omega+\omega') \breve{G}_0[\omega],
\label{G0}
\end{equation}
is ``diagonal'', where the single argument Fourier transform is similarly defined,
\begin{equation}
\label{eq-Fourier}
A[\omega] = \int_{-\infty}^{+\infty} A(t,0) e^{i\omega t} dt.
\end{equation}
(The expressions for different components of $\breve{G}_{0}[\omega]$ and $\breve{\Sigma}[\omega]$ are given in Appendix A). Using $\breve{G}_0[\omega]$, we can save one integration due to the $\delta$ function, and have
\begin{eqnarray}
\label{eq-Zsteady}
\ln {\cal Z}(\xi) &=& - \frac{1}{2} {\rm Tr}_{j,\sigma,\omega}
\ln\Bigl[ 1 - \breve{G}_0[\omega] \breve{\Sigma_L^A}[\omega, \omega'] \Bigr]
\nonumber \\
&& - \frac{i}{2 \hbar} {\rm Tr}_ {j,\sigma,\omega} \Bigl[\breve{G}[\omega,\omega'] \, \breve{{\cal F}}[\omega', \omega]\Bigr],
\end{eqnarray}
where $ \breve{G}_0[\omega] \breve{\Sigma_L^A}[\omega, \omega']$ is viewed

as a matrix indexed by $\omega$ and $\omega'$.  The trace is performed on
the frequency as well as the usual space and branch components. (The meaning of trace in frequency domain is discussed in Appendix~C). $\breve{{\cal F}}$ is given by
\begin{equation}
\label{force-matrix}
\breve{{\cal F}}[\omega,\omega'] = 
\left( \begin{array}{cc}
                             0 & 2 f[\omega]f[\omega']^T \\
                             0 & 0 
\end{array} \right).  
\end{equation}
In the next section we derive the CGF for the product initial condition using Feynman diagrammatic technique. Because of the special form of the initial density matrix the calculation for the CGF simplifies greatly in this case. 
  
\section{Product state $\rho(-\infty)$ as initial state}
In this section, we derive the CGF starting with a product initial state, i.e., the density matrix at time $t=0$ is given
by  $\rho(-\infty)=\rho_C   \otimes \rho_L \otimes \rho_R$.  Since this density matrix commutes with the projection operator $\Pi_a$, the initial projection does not play any role in this case. Working in the interaction picture with respect to the decoupled Hamiltonian ${\cal H}(-\infty)= \sum_i {\cal H}_i$, the 
interaction part of the Hamiltonian on the contour $C=[0,t_M]$ is 
\begin{eqnarray}
{\cal V}^{x}_I(\tau) &=& -f^T(\tau) u_C(\tau) + u_R(\tau) V^{RC} u_C(\tau) \nonumber \\
&&+\, u_L\bigl(\tau+\hbar x(\tau)\bigr)V^{LC}u_C(\tau).
\end{eqnarray}
In the last term for $u_L$, the argument is shifted by $\hbar x$ where 
$x^+(t) = - \xi/2$, $x^{-}(t) = \xi/2$ for $0< t < t_M$.

The density matrix remains unaffected by the transformation to the interaction picture, because it commutes with ${\cal H}(-\infty)$. The CGF can now be written as 
\begin{equation}
{\cal Z}(\xi) = {\rm Tr}\Bigl[ \rho(-\infty) T_c \,e^{-\frac{i}{\hbar} \int_C {\cal V}^{x}_I(\tau)\, d\tau}\Bigr].
\end{equation}
Expanding the exponential, we generate various terms of product of
$u_\alpha$.  These terms can be decomposed in pairs according to
Wick's theorem \cite{rammer86}. Since the system is decoupled, each type of $u$ comes
in an even number of times for a non-vanishing contributions because 
$\langle u_C\rangle = 0$, $\langle u_C u_L\rangle = 0$ and we know
\begin{equation}
-\frac{i}{\hbar} \langle T_C u_{\alpha}(\tau) u_{\alpha'}(\tau')^T \rangle_{\rho(-\infty)} = \delta_{\alpha,\alpha'} g_\alpha(\tau, \tau').
\end{equation}
We use Feynman diagrammatic technique to sum the series.  since ${\cal V}_I$ contains only two-point couplings,
the graphs are all ring type.  The combinatorial factors can be worked
out as $1/(2n)$ for a ring containing $n$ vertices.
We use a very general theorem which says $\ln {\cal Z}$ contains only 
connected graphs, and the disconnected graphs cancel exactly when we
take the logarithm. The final result can be expressed as
\begin{equation}
\ln {\cal Z}(\xi) = - \frac{1}{2} {\rm Tr}_{j,\tau} \ln \Big[ 1 - g_C \Sigma^{\rm tot} \Big] 
-\frac{i}{2\hbar}  f^T G f,  
\end{equation}
where
\begin{equation}
\Sigma^{\rm tot} = \Sigma_L(\tau+x,\tau'+x') + \Sigma_R(\tau,\tau') = \Sigma + \Sigma_L^A,
\end{equation} 
and $\Sigma$ is the total self-energy due to both the leads. $G(\tau,\tau')$ obeys the following Dyson's equation
\begin{eqnarray}
G(\tau,\tau') &=& g_C(\tau,\tau')  \\ 
&&\> + \int \int  d\tau_1 d\tau_2  g_C(\tau,\tau_1) \Sigma^{\rm tot}(\tau_1,\tau_2) G(\tau_2,\tau'). \nonumber 
\end{eqnarray}

The above expression for CGF can be written down more explicitly, by
getting rid of the vacuum diagrams. Let us define a new type of Dyson's equation
\begin{eqnarray}
\label{eq-Dyson-product}
G_0(\tau, \tau') &=& g_C(\tau,\tau') \\
&&\> + \int \!\int d\tau_1d \tau_2\, 
g_C(\tau, \tau_1) \Sigma(\tau_1,  \tau_2) G_0(\tau_2, \tau'), \nonumber
\end{eqnarray}
where $g_C$ is the contour ordered Green's function of the isolated center. (The Green's functions for an isolated single 
harmonic oscillator is given is appendix A).
This expression looks formally the same as before except that $G_0$
satisfies a Dyson equation defined on the contour from 0 to $t_M$ and
back, while $G$ is defined on the Keldysh contour from $-\infty$ to
$t_M$. Using this definition we can write
\begin{eqnarray}
1 - g_C \Sigma^{\rm tot} &=& 1 - g_C (\Sigma + \Sigma_L^A) \nonumber \\
&=& (1 - g_C \Sigma)\, (1 - G_0 \Sigma_L^A).
\end{eqnarray}
The two factors above are in matrix (and contour time) multiplication.
Using the relation between trace and determinant, 
$\ln \det(M) = {\rm Tr} \ln M$, and the fact, $\det(AB) = \det(A) \det(B)$,
we find that the two terms give two factors for ${\cal Z}$, and the factor
due to $1 - g_C \Sigma$ is exactly 1.  We have then
\begin{equation}
\label{eq-Zprod}
\ln {\cal Z}(\xi) = - \frac{1}{2} {\rm Tr}_{j,\tau} \ln \Big[ 1 - G_0 \Sigma_L^{A} \Big]
-\frac{i}{2\hbar}  f^T G f,
\end{equation}
where the $G(\tau,\tau')$ can now be expressed in terms of $G_{0}(\tau,\tau')$ as
\begin{equation}
G^{-1} = G_0^{-1} - \Sigma_L^A.
\end{equation}
which is similar in form to Eq.~(\ref{eq-Dyson-full}).

The expression for $\ln {\cal Z}(\xi)$ is consistent with the earlier result,
Eq.~(\ref{projected_state}), in the long-time limit. So we can conclude that the long-time limit is the same independent of the initial distributions. 

To compute the cumulants $\langle \langle Q^n \rangle \rangle$, we
need to take derivative with respect to $\xi$ $n$-times to $\ln {\cal Z}$.
Note that the shifted self-energy for $0 < t < t_M$ is (for all three initial conditions)
\begin{eqnarray}
\Sigma_A^t (t,t') &=& 0, \nonumber \\
\Sigma_A^{\bar t}(t,t') &=& 0, \nonumber \\
\Sigma_A^{<}(t,t') &=& \Sigma_L^{<}(t-t'-\hbar \xi) -  \Sigma_L^{<}(t-t'), \nonumber \\
\Sigma_A^{>}(t,t') &=& \Sigma_L^{>}(t-t'+\hbar \xi) -  \Sigma_L^{>}(t-t').
\label{shifted-product}
\end{eqnarray}
We note $\Sigma_L^A(\xi=0)=0$. The derivatives at $\xi=0$ can be obtained as
\begin{eqnarray}
{\partial^n \Sigma_A^{<} \over
\partial \xi^n}\Big|_{\xi=0} &=& (-\hbar)^n \Sigma_L^{<,(n)}(t-t'), \nonumber \\
{\partial^n \Sigma_A^{>} \over
\partial \xi^n}\Big|_{\xi=0} &=&  \hbar^n \Sigma_L^{>,(n)}(t-t'),
\end{eqnarray}
where the superscript $(n)$ means derivatives with respect to the argument of
the function $n$ times. In the following sections we first show the explicit expression of the CGF in the long-time limit and then discuss the steady state fluctuation theorem.

\section{Long-time limit and Steady state fluctuation theorem}
For the long-time limit calculation we can use either Eq.~(\ref{eq-Zsteady}) or Eq.~(\ref{eq-Zprod}).        
For convenience of taking the large time limit, i.e., $t_M$ large, we prefer
to set interval to $(-t_M/2, t_M/2)$. In this way, when $t_M \to \infty$,
the interval becomes the full domain and Fourier transforms to all the
Green's functions and self-energy can be performed (where the translational
invariance is restored). Applying the convolution theorem to the
trace formula in Eq.~(\ref{eq-Zprod}), we find that there is one more time integral left
with integrand independent of $t$.  This last one can be set from $-t_M/2$ to $t_M/2$, obtaining an overall factor
of $t_M$ and we have
\begin{equation}
{\rm Tr}_{(j,\tau)}(AB \cdots D) = t_M 
\int \frac{d\omega}{2\pi}{\rm Tr} \Bigl[\breve {A}(\omega)\breve{B}(\omega) \cdots \breve{D}(\omega)\Bigr].
\end{equation}
 
In the long-time limit, the shift given to the argument in $\Sigma_L^A$ depends on the branches, and the 
two arguments $(t,t')$ becomes $t-t'$ and we have 
\begin{eqnarray}
\Sigma_A^{\sigma\sigma'}(t,t') &=& \Sigma^{\sigma\sigma'}_L(t\! +\!x^\sigma\!-\! t'\!-\!x^{\sigma'}) - \Sigma^{\sigma\sigma'}_L(t\!-\!t'), \\
\Sigma_A^t &=& \Sigma_A^{\bar t} = 0,\nonumber \\
\Sigma_A^<(t) &=& \Sigma_L^{<}(t-\hbar \xi) -\Sigma_L^{<}(t),\\
 \Sigma_A^>(t) &=& \Sigma_L^{>}(t+\hbar \xi)-\Sigma_L^{>}(t). \nonumber 
\label{steady}
\end{eqnarray}
Fourier transforming the greater and lesser self-energy, we get
\begin{eqnarray}
\label{eq-a}
\Sigma_A^{>}[\omega] = \Sigma^{>}_L[\omega] \bigl(e^{-i\hbar \omega \xi} - 1 \bigr) = a, \\
\label{eq-b}
\Sigma_A^{<}[\omega] = \Sigma^{<}_L[\omega] \bigl(e^{i\hbar \omega \xi} - 1 \bigr) = b.
\end{eqnarray}
We note that $\Sigma_L^A$ is supposed to depend on both $\xi$ and $\lambda$.
However in the long-time limit, the $\lambda$ dependence drops out which makes the steady state result independent of the initial distribution.  

Finally, we can express the generating function as
\begin{eqnarray}
\ln {\cal Z}(\xi) &= & - t_M \int \frac{d\omega}{4\pi} {\rm Tr} \ln \Bigl[1 - \breve{G}_0 [\omega]\breve{\Sigma}_L^A [\omega]\Bigr] \nonumber \\
&& \quad - \frac{i}{\hbar}  \int \frac{d\omega}{4\pi}
{\rm Tr} \Bigl[ \breve{G}[\omega] \breve{\cal F}[\omega,-\omega]\Bigr],
\label{eq-lnZxi-1}
\end{eqnarray}
where $\breve{G}[\omega]$ is obtained by solving the Dyson equation
in frequency domain and in the long-time obeys time-translational invariance. So the full CGF can be written as the sum of contributions due to driving force and due to temperature difference between the leads, i.e.,
\begin{equation}
\ln {\cal Z}(\xi) = \ln {\cal Z}^{s}(\xi) + \ln {\cal Z}^{d}(\xi). 
\end{equation}
In the following and subsequent sections we discuss about ${{\cal Z}^{s}(\xi)}$ and we will return to ${{\cal Z}^{d}(\xi)}$ in Sec. XI. 

In order to obtain the explicit expression for $\ln {\cal Z}^{s}(\xi)$ we need to compute the matrix product
\begin{eqnarray}
\breve{G}_0[\omega] \breve{\Sigma}_L^A[\omega]  &=& 
\frac{1}{2} \left(  \begin{array}{cc}
G_0^r & G_0^{K} \\
0 & G_0^a  
\end{array}
\right)
\left(  \begin{array}{cc}
a-b & a+b \\
-(a+b) & b-a  
\end{array}
\right).
\end{eqnarray}
To simplify the expression, we rewrite the term ${\rm Tr} \ln (1-M)$ as
a determinant and use the formula (assuming A to be an invertible matrix)  
\begin{equation}
{\rm det} \left( \!\!\begin{array}{cc}
A & B \\
C & D \end{array}\!\!
\right) = {\rm det}(A) \det(D - C A^{-1}B)
\end{equation}
to reduce the dimensions of the determinant matrix by half. 
The steady state solution for ${{\cal Z}^{s}(\xi)}$ is given by
\begin{eqnarray}
\ln {\cal Z}^{s}(\xi)&=&-t_M \int \frac{d\omega}{4\pi} \,\ln \det \Bigl\{ I - G_{0}^r \Gamma_L 
G_{0}^a \Gamma_R \Big[  (e^{i\xi \hbar \omega}\! -\! 1) f_L \nonumber \\ 
&&\!\!\!\!\!\!\!\!\!\!\!\!+ ( e^{-i\xi\hbar \omega} \!-\! 1) f_R + (
e^{i\xi \hbar \omega} \!+\! e^{-i\xi\hbar\omega} \!-\!2 ) f_L f_R \Big]\Bigr\}.\qquad 
\label{eq-steady1}
\end{eqnarray}
with $f_{\alpha}=1/(e^{\beta_{\alpha} \hbar \omega} - 1)$, $\alpha=L,R,$ the Bose-Einstein distribution function and $\Gamma_{\alpha}[\omega]=i\big(\Sigma_{\alpha}^{r}[\omega]-\Sigma_{\alpha}^{a}[\omega]\big)$. If we consider the full system as a one-dimensional linear chain, then because of the special form of $\Gamma_{\alpha}$ matrices (only one entry of the $\Gamma$ matrices are non-zero) it can be easily shown that
\begin{equation}
\det [I-\bigl(G_{0}^r \Gamma_L G_{0}^a \Gamma_R \bigr) \Xi(\xi)]=1-{\cal T}[\omega]\Xi(\xi)
\end{equation}
where $\Xi(\xi)$ is any arbitrary function of $\xi$ and ${\cal T}[\omega]={\rm Tr}(G_{0}^r \Gamma_L G_{0}^a \Gamma_R)$ is known as the transmission function and is given by the Caroli formula \cite{Caroli,WangJS-europhysJb-2008}. The generating function ${{\cal Z}^{s}(\xi)}$ in the steady state obeys the following symmetry  
\begin{equation}
{{\cal Z}^{s}}(\xi)={{\cal Z}^{s}}\big(-\xi + i\,{\cal A}\big),
\end{equation}
where ${\cal A}=\beta_R-\beta_L$ is known as thermodynamic affinity. This relation is also known as Gallavotti-Cohen (GC) symmetry \cite{noneq-fluct}. The immediate consequence of this symmetry is that the probability distribution for heat transferred $Q_{L}$ which is given by the Fourier transform of the CGF, i.e., $P(Q_L)=\frac{1}{2 \pi} \int_{-\infty}^{\infty} d\xi \, {\cal Z}(\xi) \, e^{-i \xi Q_L}$ obeys the following relation in the large $t_M$ limit,
\begin{equation}
P_{t_M}(Q_L)=e^{{\cal A}Q_L}\,P_{t_M}(-Q_L).
\end{equation} 
This relation is known as the steady state fluctuation theorem and was first derived by Saito and Dhar \cite{Saito-Dhar-2007} in the phononic case. This theorem quantifies the ratio of positive and negative heat flux and second law violation. 

The cumulants $\langle \langle {Q}^{n} \rangle \rangle $ can be obtained by taking derivative of $\ln {\cal Z}^{s}(\xi)$ with respect to $i\xi$ and setting $\xi=0$. The first cumulant is given by 
\begin{equation}
\frac{\langle \langle Q \rangle \rangle}{t_M} = \int_{-\infty}^{\infty} \frac{d\omega}{4 \pi} \, \hbar\, \omega \,{\cal T}(\omega)(f_L-f_R),
\end{equation}
which is known as the Landauer-like formula in thermal transport. Similarly the second cumulant $\langle \langle {Q}^{2} \rangle \rangle= \langle {Q}^{2} \rangle - \langle {Q}\rangle ^{2}$, which describes the fluctuation of the heat transferred, can be written as \cite{Saito-Dhar-2007, Huag, Buttiker},
\begin{eqnarray}
\frac{\langle \langle Q^{2} \rangle \rangle}{t_M} &=&\int_{-\infty}^{\infty} \frac{d\omega}{4 \pi} \, (\hbar \omega)^{2} \Bigl \{{\cal T}^{2}(\omega)\, (f_L-f_R)^{2} \nonumber \\
&&+ {\cal T}(\omega) \, (f_L+f_R+2\,f_L f_R ) \Bigr\}.
\end{eqnarray}
Our formalism can be easily generalized for multiple heat baths and for $N$ leads connected with the center $C$, we can generalize the above formula as
\begin{eqnarray}
&&\ln {\cal Z}^{s}(\xi)=\!-\!t_M \!\int \frac{d\omega}{4\pi} \,\ln \det \Bigl\{ I \!-\!\sum_{m} G_{0}^r \Gamma_L G_{0}^a \Gamma_m \Big[(e^{i\xi \hbar \omega}\! -\! 1) \times \nonumber \\
&& f_L\!\!+ ( e^{-i\xi\hbar \omega} \!-\! 1) f_m \!+\! (\!
e^{i\xi \hbar \omega} \!\!+\! e^{-i\xi\hbar\omega}\! \!-\!2 )\! f_L \!f_m \Big]\Bigr\}.\qquad .
\end{eqnarray}
In the following section we will discuss the how to numerically calculate the CGF in the transient case for projected density matrix $\rho'(0)$. We also discuss about solving the Dyson equation given in Eq.~(\ref{eq-Dyson-product}).

\section{Transient Region}
The central quantity to calculate the CGF numerically is the shifted self-energy $\Sigma_L^{A}$ which is given by 
\begin{equation}
\Sigma_L^A(\tau,\tau') =\Sigma_L\bigl(\tau + \hbar x(\tau), \tau'+\hbar x(\tau') \bigr)- \Sigma_L\big(\tau,\tau'\big).
\end{equation}
Here $\tau$ is a contour variable which runs over Keldysh contour $K=(-\infty,\infty)$ and back, for the initial conditions $\rho(0)$ and $\rho'(0)$, whereas for $\rho(-\infty)$, $\tau$ runs over the contour $C=[0,t_M]$ (see Fig.~1). The contour function $x(\tau)$ is 0 whenever $t < 0$ or $t>t_M$, and for $0 < t < t_M$, $x^{+}(t) = -\xi/2 - \lambda$, and $x^{-}(t) = \xi/2 - \lambda$. 
Depending on the values of $t$, $t'$, and $\lambda$ ($\lambda \rightarrow 0$  and $\lambda \rightarrow \infty$ corresponds to steady state initial state and {\it projected} initial state, respectively) $\Sigma_L^A$ will have different functional form. If $ 0 < t,t' < t_M$ then $\Sigma_L^A$'s are given by Eq.~(\ref{shifted-product}). This is the region which dominates in the long-time limit and gives steady state result. If both $t$ and $t'$ lies outside the measurement time, i.e., $t,t'<0$ or $t,t'>t_M$ then $\Sigma_L^A$ is zero. 

The main computational task for a numerical evaluation of the cumulants
is to compute the matrix series $-\ln(1-M) = M + \frac{1}{2} M^2 + \cdots$.
It can be seen due to the nature of $\Sigma_L^A$ that for the product initial
state, exact $n$ terms upto $M^n$ is required for the $n$-th culumants, as the infinite series terminates due to $\Sigma_L^A(\xi=0) = 0$.
Numerically, we also observed for the projected state $\rho'(0)$, exactly
$3n$ terms is required (although we don't have a proof) if calculation is
performed in time domain. 

The computation can be performed in time as well as in the frequency domain. However for projected and steady state initial condition since $G_{0}[\omega]$ is time translational invariant it is advantageous to work in the frequency domain. But for the product state there is no such preference as $G_{0}$ in Eq.~(\ref{eq-Dyson-product}) is not time translational invariant and one has to solve it numerically. 

In the following we first discuss how to calculate $\Sigma_{L}^{A}[\omega,\omega']$ for projected initial state, defined in Eq.~(\ref{eq-Zsteady}) and then we will discuss how to solve the Dyson equation for the product initial condition case, given in Eq.~(\ref{eq-Dyson-product}).

\subsection{calculation of $\Sigma_L^{A}(\omega,\omega')$ }
To calculate $\Sigma_L^{A}(\omega,\omega')$ for projected initial state $\rho'(0)$ we define two types of theta functions $\theta_{1}(t,t')$ and $\theta_{2}(t,t')$.
$\theta_{1}(t,t')$ is non-zero when 
\begin{equation}
0 \leq t \leq t_M, \,\,{\rm and} \quad t' \leq 0 \quad {\rm or} \quad t'\geq t_M,
\end{equation}
or 
\begin{equation}
0 \leq t' \leq t_M, \,\,{\rm and} \quad t \leq 0 \quad {\rm or} \quad t \geq t_M, 
\end{equation}
and $\theta_{2}(t,t')$ is non-zero only in the regime where $ 0 \le t,t' \le t_M$ . For the regions where $\theta_1(t,t')$ is non-zero the expression for $\Sigma_L^{A}$ after taking the limit $\lambda \rightarrow \infty$ is, (assuming all correlation functions decays to zero as $t \rightarrow \pm \infty$)
\begin{equation}
\Sigma_{A}^{t,\bar{t},<,>}(t,t')=-\Sigma_{L}^{t,\bar{t},<,>}(t-t').
\end{equation}
So using theta functions we may write $\Sigma_L^{A}(t,t')$ in the full t,t' domain as 
\begin{eqnarray}
\Sigma_{A}^{t,\bar{t}}(t,t')&=&-\theta_{1}(t,t')\Sigma_{L}^{t,\bar{t}}(t-t') \nonumber \\
\Sigma_{A}^{<}(t,t')&=&-\theta_{1}(t,t')\Sigma_{L}^{<}(t-t')+ \theta_2 (t,t') \times \nonumber \\
&&\>\big[\Sigma_{L}^{<}(t-t'-\hbar \xi) -  \Sigma_{L}^{<}(t-t')\big] \nonumber \\
\Sigma_{A}^{>}(t,t')&=&-\theta_{1}(t,t')\Sigma_{L}^{>}(t-t')+ \theta_2 (t,t') \times \nonumber \\
&&\>\big[\Sigma_{L}^{>}(t-t'+\hbar \xi) -  \Sigma_{L}^{>}(t-t')\big]
\end{eqnarray}
By doing Fourier transform it can be easily shown that
\begin{equation}
\Sigma_{A}^{t,\bar{t}}[\omega,\omega']=-\int_{-\infty}^{\infty}\!\!\! \frac{d\omega_{c}}{2 \pi} \, \theta_{1}\bigl[\omega\!-\!\omega_{c},\omega'\!+\!\omega_{c}\bigr]\Sigma_{L}^{t,\bar{t}}(\omega_{c})
\end{equation}
and 
\begin{eqnarray}
\Sigma_{A}^{>,<}[\omega,\omega']&\!=\!& \!-\!\int_{-\infty}^{\infty} \!\!\!\frac{d\omega_{c}}{2 \pi}\theta_{1}\bigl[\omega\!-\!\omega_{c},\omega'\!+\!\omega_{c}\bigr]\Sigma_{L}^{>,<}(\omega_{c}) \\
\!+\!\int_{-\infty}^{\infty}\! \frac{d\omega_{c}}{2 \pi}\!\!\!&&\!\theta_{2}\bigl[\omega\!-\!\omega_{c},\omega'\!+\!\omega_{c}\bigr]\Sigma_{L}^{>,<}(\omega_{c}) 
(e^{i \omega_{c}\eta \xi}\!-\!1), \nonumber
\end{eqnarray}
where $\eta=\pm 1$. The positive sign is for $\Sigma_{A}^{<}$ and negative sign for  $\Sigma_{A}^{>}$.

The theta functions are now given by
\begin{eqnarray}
\theta_{1}(\omega_{a},\omega_{b})&=&f(\omega_{a}).g(\omega_{b})+f(\omega_{b}).g(\omega_{a}), \nonumber \\
\theta_{2}(\omega_{a},\omega_{b})&=&f(\omega_{a}).f(\omega_{b}),
\end{eqnarray}
where 
\begin{eqnarray}
f(\omega)&=&\frac{e^{i \omega t_{M}}-1}{i \omega}, \nonumber \\
g(\omega)&=& \frac{1}{i \omega + \epsilon} -\frac{e^{i \omega t_{M}-\eta t_{M}}}{i \omega -\epsilon},
\end{eqnarray}
with $\epsilon \rightarrow 0^{+}$. The theta functions are of immense importance which carries all information about the measurement time $t_{M}$.

In the limit $t_{M} \rightarrow \infty$, the region  $ 0 \le t,t' \le t_M$ dominates and corresponding theta function, i.e., $\theta_2(\omega,\omega')$ reduces to 
\begin{equation}
\theta_{2}(\omega-\omega_{c},\omega'+\omega_{c}) \approx \delta(\omega-\omega_{c}) \delta(\omega'+\omega_{c}),
\end{equation}
and is responsible for obtaining the steady state result.

To calculate all the cumulants we only need to take derivative of $\Sigma_{A}(\omega,\omega')$ with respect to $i \xi$ since $G_{0}$ does not have any $\xi$ dependence. Also $\Sigma^{A}$ has $\xi$ dependence only for $ 0 \le t,t' \le t_M$ and hence the derivatives are given by 
\begin{eqnarray}
\frac{\partial^{n}\Sigma_{A}^{>,<}}{{\partial}(i \xi)^{n}}[\omega,\omega']&=&\int_{-\infty}^{\infty} \frac{d\omega_{c}}{2 \pi} (\eta \hbar \omega_{c})^{n} \theta_{2}\bigl[\omega-\omega_{c},\omega'+\omega_{c}\bigr]\nonumber \\
&&\Sigma_{L}^{>,<}(\omega_{c}) e^{i \omega_{c}\eta \xi}.
\end{eqnarray}  
Here $n$ refers to the order of the derivative. 

\subsection{Dyson equation on contour C}

Let us now discuss about solving the Dyson's equation for $G_{0}$ given in Eq.~(\ref{eq-Dyson-product}) for product initial state $\rho(-\infty)$. In order to compute the matrix $\breve{G}_0(t,t')$ we have to calculate two components $G_{0}^{r}$ and $G_{0}^{K}$ which are written in the integral form by applying Langreth's rule \cite{Huag,rammer86}
\begin{eqnarray}
G_{0}^{r}(t,t')&=&g_{C}^{r}(t\!-\!t')  \\
&&+\!\!\int_{0}^{t_M}\!\!dt_1\!\!\!\int_{0}^{t_M}\!\!dt_2 \, g_{C}^{r}(t\!-\!t_1)\,\Sigma^{r}(t_1\!-\!t_2)G_{0}^{r}(t_2,t'), \nonumber 
\end{eqnarray}
and 
\begin{eqnarray}
&&G_{0}^{K}(t,t')=g_{C}^{K}(t\!-\!t') \\
&&+\!\!\int_{0}^{t_M}\!\!dt_1\!\!\!\int_{0}^{t_M}\!\!dt_2 \, g_{C}^{r}(t\!-\!t_1)\,\Sigma^{r}(t_1\!-\!t_2) G_{0}^{K}(t_2,t') \nonumber \\
&&+\!\!\int_{0}^{t_M}\!\!dt_1\!\!\!\int_{0}^{t_M}\!\!dt_2 \, g_{C}^{r}(t\!-\!t_1) \Sigma^{K}(t_1\!-\!t_2) G_{0}^{a}(t_2,t')\nonumber \\
&&+\!\!\int_{0}^{t_M}\!\!dt_1\!\!\!\int_{0}^{t_M}\!\!dt_2 \, g_{C}^{K}(t\!-\!t_1) \Sigma^{a}(t_1\!-\!t_2)G_{0}^{a}(t_2,t'). \nonumber 
\end{eqnarray}

Note that the argument for center Green's function $g_{C}$ and lead self-energy $\Sigma$ are written as time difference $t-t'$ because they are Green's functions for isolated center part and leads respectively and hence are calculated at equilibrium. The analytical expressions for $\Sigma$ and $g_{C}$ are known in frequency domain and are given in Appendix~A. To determine their time-dependence we numerically calculate their inverse Fourier transforms using trapezoidal rule \cite{NR}. Then in order to solve above equations for any $t_M$  we discretize the time variable into $N$ total intervals of incremental length $\Delta t=t_M/N$ and thus converting the integral into a sum. After discretization, the above equations can be written in the matrix form which are indexed by space $j$ and discrete time $t$, as 
\begin{eqnarray}
\tilde{G}_{0}^{r}&=& \tilde{g}_C^{r} + \tilde{g}_C^{r} \tilde{\Sigma}^{r} \tilde{G}_{0}^{r}, \nonumber \\
\tilde{G}_{0}^{K}&=& \tilde{G}_{0}^{r}\tilde{\Sigma}^{K}\tilde{G}_{0}^{a}+ ( I + \tilde{G}_{0}^{r} \tilde{\Sigma}^{r}) \tilde{g}_C^{K} (I+\tilde{\Sigma}^{a} \tilde{G}_{0}^{a}).
\end{eqnarray}
So $\tilde{G}_{0}^{r}$ can be obtained by doing an inverse of the matrix $(I-\tilde{g}_{C}^{r} \tilde{\Sigma}^{r})$ and then multiplying by $\tilde{g}_{C}^{r}$. $\tilde{G}_{0}^{r}$ in this case also obeys time-translational invariance, so it can also be obtained by direct inverse Fourier transform. $\tilde{G}_{0}^{a}$ can be obtained by taking transpose of $\tilde{G}_{0}^{r}$. Once $\tilde{G}_{0}^{r}$ and $\tilde{G}_{0}^{a}$ are obtained we use the second equation to calculate $\tilde{G}_{0}^{K}$ which is simply multiplying matrices.

Similarly $\Sigma_L^{A}$ in Eq.~(\ref{shifted-product}) are obtained by doing inverse Fourier transforms of the lead self-energy. We follow the same steps independently to calculate the cumulants for $Q_R$. 

\begin{figure}
\includegraphics[width=\columnwidth]{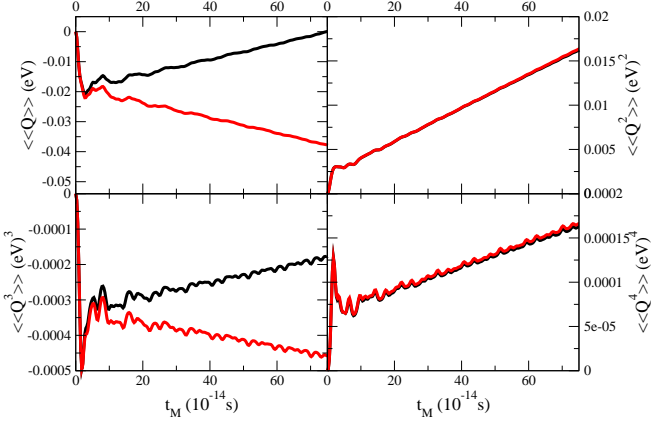}
\caption{(Color online) The cumulants $\langle \langle {Q}_{L}^{n} \rangle \rangle$ and $\langle \langle {Q}_{R}^{n} \rangle \rangle$ for $n$=1, 2, 3, and 4 for one-dimensional linear chain connected with Rubin baths, for the projected initial state $\rho'(0)$. The black and red curves corresponds to $\langle \langle {Q}_{L}^{n} \rangle \rangle$ and  $\langle \langle {Q}_{R}^{n} \rangle \rangle$ respectively. The temperatures of the left and the right lead are 310 K and 290 K, respectively. The center (C) consists of one particle.}
\end{figure}

\begin{figure}
\includegraphics[width=\columnwidth]{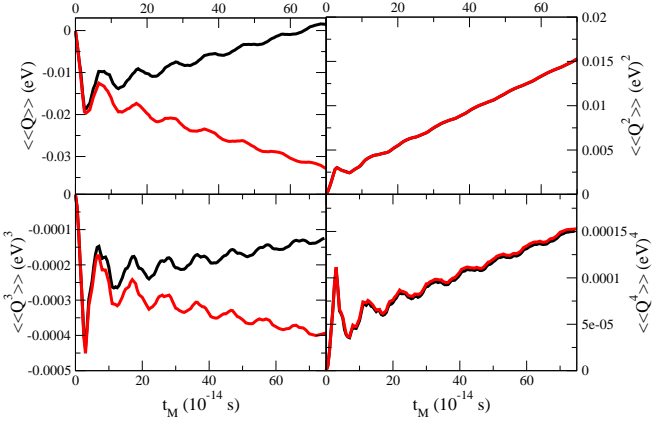}
\caption{(Color online) Same as in Fig.~2 except for product initial state $\rho(-\infty)$. The temperatures of the left, the center and the right lead are 310 K, 300 K and 290 K, respectively.}
\end{figure}

\begin{figure}
\includegraphics[width=\columnwidth]{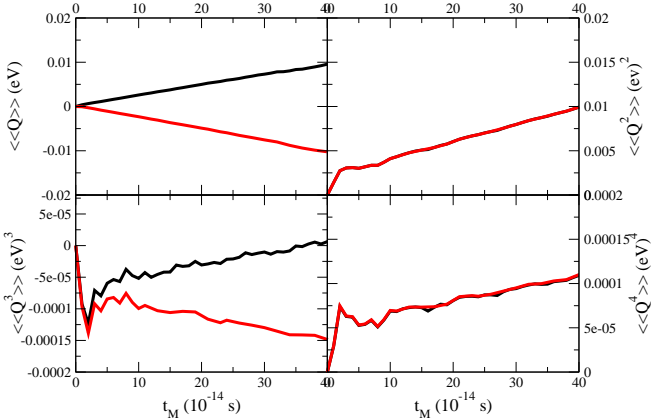}
\caption{(Color online) Same as in Fig.~2 except for steady state initial state $\rho(0)$.}
\end{figure}

\section{Numerical results}
We now present some numerical results. In Fig.~2 and 3, we show the results for first four cumulants for both ${\cal Q}_L$ and ${\cal Q}_R$ (measurement is on the right lead) for 1D linear chain connected with Rubin baths, starting with the projected initial state $\rho'(0)$ and product state $\rho(-\infty)$ respectively. 
Rubin baths \cite{Rubin, Weiss} mean in our case a uniform linear
chain with spring constant $K$ and a small onsite $K_0$ for all the atoms.
Only one atom is considered as the center.  The atoms of the left and right side
of the center are considered baths.  We use $K=1$ eV/(u\AA$^2)$  and the onsite potential $K_0=0.1$ eV/(u\AA$^2)$ in all our calculations.
First of all, cumulants greater than two are nonzero, which confirms that the distribution for $P(Q_L)$ or $P(Q_R)$ is not Gaussian.  The generic features are almost the same in both the cases. However the fluctuations are larger for the product initial state $\rho(-\infty)$ as this state corresponds to the sudden switch on of the couplings between the leads and the center and hence the state is far away from the correct steady state distribution. On the contrary, for the initial state $\rho'(0)$ the fluctuations are relatively small. For $\rho'(0)$ due to the effect of the measurement, at starting time energy goes into the leads, which is quite surprising. But for $\rho(-\infty)$ although initial measurement do not play any role, energy still goes into the leads. This can also be shown analytically (see Appendix B). At the starting time the behavior of both ${\cal Q}_L$ and ${\cal Q}_R$ are very similar and can be understood since both the left and right leads are identical and the effect of temperature difference is not present. However at longer times the odd cumulants starts differing and finally grows linearly with time $t_M$ and agrees with the corresponding long-time predictions.

In Fig.~4 we show the results for the steady state initial condition, i.e., $\rho(0)$ which can be obtained by mapping the projection operators as identity operator, i.e., taking the limit $\lambda \rightarrow 0$. So in this case measurement effect is ignored and the dynamics starts with the actual steady state for the full system. The first cumulant increases linearly from the starting, $\langle Q \rangle = t I$ and the slope gives the correct prediction with the Landauer-like formula. However, high order cumulants still have transient behavior. In this case the whole system achieve steady state much faster compared with the other two cases.

\section{correlation between left and right lead heat}
\subsection{Product initial state}
In this section, we derive the CGF for the joint probability distribution $P(Q_L,Q_R)$ for the product initial state $\rho(-\infty)$. In order to calculate the CGF we need to measure both ${\cal H}_L$ and ${\cal H}_R$ at time 0 and at time $t_M$. Since the Hamiltonians for the left and the right lead commute at the same instance of time i.e., $\big[{\cal H}_L, {\cal H}_R \big]=0$, such type of measurements are allowed in quantum mechanics and also Nelson's theorem \cite{Nelson} gurentee's that $P(Q_L,Q_R)$ is a well-defined probability distribution. The immediate consequence of deriving such CGF is that, the correlations between the left and the right lead heat can be obtained and it is also possible to calculate the CGF for total entropy flow (defined below) to the reservoirs. To calculate the CGF we need two counting fields $\xi_L$ and $\xi_R$ and the CGF in this case can be written down as \cite{Esposito-review-2009}
\begin{equation}
\mathcal {Z}(\xi_L,\xi_R)=\langle e^{i\,\xi_L\,{\cal H}_L+i\,\xi_R\,{\cal H}_R}\,\, e^{-i\,\xi_L\,{\cal H}^{H}_L(t)-i\,\xi_R\,{\cal H}^{H}_R(t)} \rangle',
\end{equation}
where the average is defined as 
\begin{equation}
\langle \cdots \rangle'=\sum_{a,c}\Pi^{L}_a \, \Pi^{R}_c \, \rho(0)\, \Pi^{L}_a \,\Pi^{R}_c.
\end{equation}
$\Pi^{L}_a$ and $\Pi^{R}_c$ are the projectors onto the eigenstates of ${\cal H}_L$ and ${\cal H}_R$ with eigenvalues $a$ and $c$ respectively, corresponding to the measurements at $t=0$. Here we will consider only the product state $\rho(-\infty)$, then initial projections $\Pi^{L}_a$ and $\pi^{R}_c$ do not play any role. We can proceed as before and finally the CGF can be written down as
\begin{equation}
 \ln {\cal Z}(\xi_L,\xi_R) = \sum_{k=1}^\infty \frac{1}{2k}
{\rm Tr}_{(j,\tau)} \Bigl[\big( G_{0} (\Sigma_L^{A}+\Sigma_R^A) \big)^k \Bigr],
\label{eq-lead-lead}
\end{equation}
i.e., in this case we need to shift the contour-time arguments for both left and right lead self-energies. In the long-time limit ${\cal Z}(\xi_L,\xi_R)$ becomes a function of difference of counting field $\xi_L$ and $\xi_R$, i.e., $\xi_L-\xi_R$. The explicit expression for the CGF in the long-time limit is 
\begin{eqnarray}
&&\ln {\cal Z}(\xi_L-\xi_R)=-t_M \int\frac{d\omega}{4\pi} \ln \det \Bigl\{ I - G_{0}^r \Gamma_L 
G_{0}^a \Gamma_R \nonumber \\ &&\big[(e^{i(\xi_L-\xi_R) \hbar \omega}\! -\! 1) f_L  
+( e^{-i(\xi_L-\xi_R)\hbar \omega} \!-\! 1) f_R  \nonumber \\
&&+(e^{i(\xi_L-\xi_R) \hbar \omega} \!+\! e^{-i(\xi_L-\xi_R)\hbar\omega} \!-\!2 ) f_L f_R \big]\Bigr\}.\qquad
\label{eq-lnZxi}
\end{eqnarray}
where $G_{0}$ obeys the same type of Dyson equation as in Eq.~(\ref{eq-Dyson-product}). This CGF in the steady state obeys the same type of GC fluctuation symmetry, which in this case is given by 
\begin{equation}
{\cal Z}(\xi_L-\xi_R)={\cal Z}(-\xi_L+\xi_R +i {\cal A}).
\end{equation}
Now performing Fourier transform of the CGF, the joint probability distribution is given by $P(Q_L,Q_R)=P(Q_L)\,\delta(Q_L+Q_R)$. The appearance of the delta function is a consequence of the energy conservation in the steady state, i.e., $I_L=-I_R$. In the steady state knowing probability distribution either for ${\cal Q}_L$ or ${\cal Q}_R$ is sufficient to know the joint probability distribution.

The cumulants can be obtained from the CGF by taking derivatives with respect to both $\xi_L$ and $\xi_R$, i.e.,$\langle \langle {Q}_L^{n} {Q}_R^{m} \rangle \rangle =\partial^{n+m} \ln {\cal Z}/\partial (i\xi_L)^n \partial (i\xi_R)^m,$ substituting $\xi_L=\xi_R=0.$ In the steady state the cumulants obey $\langle \langle {Q}_L^{n} {Q}_R^{m} \rangle \rangle= (-1)^{m} \langle \langle Q_L^{m+n} \rangle \rangle = (-1)^{n} \langle \langle Q_R^{m+n} \rangle \rangle$. The first cumulant give us the left and right lead correlation $\langle \langle {Q}_{L} {Q}_{R} \rangle \rangle=\langle {Q}_{L} {Q}_{R} \rangle - \langle {Q}_{L} \rangle \langle {Q}_{R} \rangle$ and in the steady state is equal to $-\langle \langle {Q}_L^{2} \rangle \rangle$.

In Fig.~5 we plot the first three cumulants for one dimensional linear chain connected with Rubin bath where the center consists of only one atom. Initially the cumulant $\langle \langle {Q}_{L} {Q}_{R} \rangle \rangle $ is positively correlated as both $Q_L$ and $Q_R$ are negative, however in the longer time since $Q_L=-Q_R$ the correlation becomes negative. We also give plots for  $\langle \langle {Q}_L^{2} {Q}_R \rangle \rangle$ (black online) and  $\langle \langle {Q}_R^{2} {Q}_L \rangle \rangle$ (red online) which are in the long-time limit 
negative and positively correlated respectively and match with the long-time predictions.

\begin{figure}
\includegraphics[width=\columnwidth]{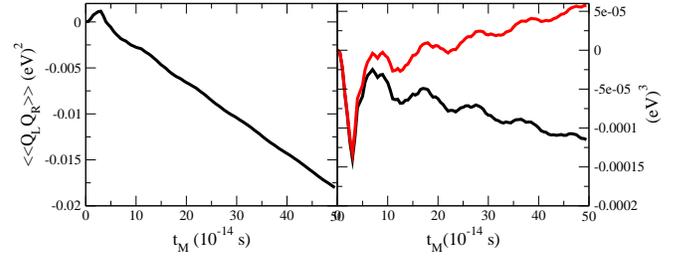}
\caption{(Color online) First three cumulants of the correlations between left and right lead heat flux for one dimensional linear chain connected with Rubin baths, starting with product initial state $\rho(-\infty)$. The left graph corresponds to $\langle \langle {Q}_{L} {Q}_{R} \rangle \rangle$ and the right graph corresponds to cumulants $\langle \langle {Q}_L^{2} {Q}_R \rangle \rangle$ (Black curve) and  $\langle \langle {Q}_R^{2} {Q}_L \rangle \rangle$ (Red curve). The left, center and right lead temperatures are 310 K, 290 K and 300 K respectively. The center (C) consists of one particle.}
\end{figure}

\begin{figure}
\includegraphics[width=\columnwidth]{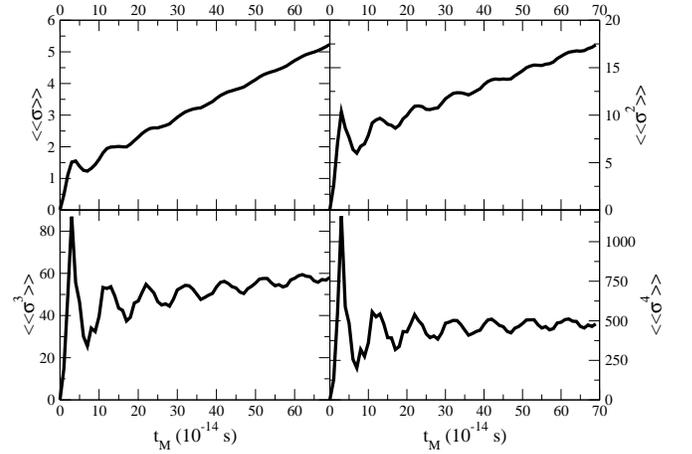}
\caption{The cumulants of entropy production $\langle \langle \sigma^{n} \rangle \rangle$ for $n$=1, 2, 3, 4 for one dimension linear chain connected with Rubin baths, for product initial state $\rho(-\infty)$. The left, center and right lead temperatures are 510 K, 400 K, and 290 K respectively. The center (C) consists of one particle.}
\end{figure}

\subsection{Entropy flow to the reservoir}
From the two parameter ($\xi_L,\xi_R$) CGF one can also obtain the total entropy that flows into the leads. The total entropy flow to the reservoirs can be defined as \cite{ep1,ep2}
\begin{equation}
{\cal \sigma}=-\beta_{L}{\cal Q}_{L} -\beta_{R}{\cal Q}_{R}.
\end{equation}
In order to calculate this CGF we just make the substitutions $\xi_L \rightarrow -\beta_L  \mu$ and $ \xi_R \rightarrow - \beta_R  \mu$ in Eq.~(\ref{eq-lead-lead}).
In the long-time limit the expression for entropy-production is similar to $\ln {\cal Z}(\xi_L,\xi_R)$ with $\xi_L-\xi_R$ replaced by ${\cal A}$ and becomes an explicit function of thermodynamic affinity $\beta_R-\beta_L$ \cite{fluct-theorems}. The CGF in this case satisfies the following symmetry
\begin{equation}
{\cal Z}(\mu)={\cal Z}(-\mu + i)
\end{equation}
In Fig.~6 we give results for the first four cumulants of the entropy flow. All cumulants are positive and in the long-time limit give correct predictions.

\section{Long-time result for $\ln {\cal Z}^d(\xi)$}

In this section we derive the explicit expression for the long-time limit of the CGF $\ln {\cal Z}^d(\xi)$ which is given by (Eq.~\ref{eq-lnZxi-1})
\begin{equation}
\ln {\cal Z}^d(\xi)=- \frac{i}{\hbar}  \int \frac{d\omega}{4\pi}
{\rm Tr} \bigl[ \breve{G}[\omega] \breve{{\cal F}}[\omega,-\omega]\bigr].
\end{equation}
where $G[\omega]$ obeys the Dyson equation given in Eq.~(\ref{eq-Dyson-full}). It is possible to write down  $\breve{G}[\omega]$ in terms of $\breve{G}_{0}$ and $\breve{\Sigma}_L^{A}$ as $\breve{G}[\omega]=\big(I-\breve{G}_{0}\breve{\Sigma}^{A}_L\big)^{-1} \breve{G}_{0}[\omega]$. This equation can be solved analytically. Next we assume that the product of $f(t)$ and $f(t')$ is a time-translationally invariant function, i.e., $f(t)f^{T}(t')=F(t-t')$ in order to get rid of $t+t'$ dependence term. In the Fourier domain this means $f[\omega]f^{T}[\omega']=2\pi F[\omega] \delta(\omega+\omega')$. So from Eq.~(\ref{force-matrix}) the matrix element ${\cal F}_{12}$ is given by $\breve{{\cal F}}[\omega,-\omega]_{12} \propto \delta(0) F[\omega]$. We write $\delta(0)=t_M/2\pi$. Using these results the CGF can be expressed as

\begin{equation}
 \ln {\cal Z}^{d}(\xi)= {i t_M}  \int \frac{d\omega}{4\pi \hbar} \, \frac{1} {{\cal N}(\xi)}{\rm Tr} \Big[G_{0}^{r}[\omega] (a+b) G_{0}^{a}[\omega] F[\omega] \Big],
\end{equation}
where $a$ and $b$ are defined in Eq.~(\ref{eq-a}) and Eq.~(\ref{eq-b}). Using the expressions for the self-energy the CGF reduces to 
\begin{equation}
 \ln {\cal Z}^{d}(\xi)=\!\!\int \!\frac{d\omega}{4 \pi \hbar}\,\frac{\cal{K}(\xi)}{{\cal N}(\xi)}\, {\rm Tr} \Big[G_{0}^{r}[\omega] \Gamma_{L}[\omega] G_{0}^{a}[\omega]F[\omega] \Big],
\label{eq-driven-CGF}
\end{equation}
with 
\begin{equation}
{\cal{K}}(\xi)=(e^{-i\xi\hbar \omega}\!-\!1 )+f_L(e^{i\xi\hbar\omega}\!+\! e^{-i\xi\hbar \omega}\!-\!2),
\end{equation} 
and 
\begin{eqnarray}
{\cal {N}}(\xi)&=& {\rm det} \Big[I -\big(G_{0}^r \Gamma_L G_{0}^a \Gamma_R\big) \Bigl \{ (f_L e^{i\xi \hbar \omega}\!-\!1) + f_R \nonumber \\ 
&&( e^{-i\xi\hbar \omega}\!-\!1) + (e^{i\xi \hbar \omega}\!+\! e^{-i\xi\hbar\omega}\!-\!2 ) f_L f_R  \Bigr\}.
\end{eqnarray}
It is important to note that ${\cal K}(\xi)$ depends only on left lead temperature and satisfies the symmetry ${\cal K}(\xi)={\cal K}(-\xi-i\beta_L)$. So we can immediately write ${\cal Z}^{d}(-i\beta_L)=1$ and this relation is completely independent of the information about the right lead. If we consider the two leads at the same temperature ($\beta_L=\beta_R=\beta$), this form of symmetry is then closely related to the Jarzynski equality (JE) \cite{JE,talkner2008} and ${\cal Z}^{d}(-i\beta)=1$ is one special form of JE. However since ${\cal N}(\xi)$ does not satisfy this particular symmetry of $\xi$ at thermal equilibrium (it obeys the GC summery when the leads are at different temperatures) and the CGF $\ln{\cal Z}^{d}(\xi)$ doesn't satisfy any such symmetry relation and hence JE is not satisfied. This does not violate JE as our definition of ${\cal Z}^{d}(\xi)$ is different from the one used to derive JE.
  
Let us now come back to the general scenario with leads at different temperatures and give the explicit expression of first and second cumulant by taking derivative of $\ln {\cal Z}^{d}(\xi)$ with respect to $i\xi$. 

The first cumulant or moment is given by \cite{driven-bijay}
\begin{equation}
\frac{\langle \langle Q_{d} \rangle \rangle}{t_M} = - \int \frac{d\omega}{4 \pi} \, \omega \, {\cal S}[\omega],
\end{equation}
where we define ${\cal S}[\omega]$ as the transmission function for the driven case and is given by
\begin{equation}
{\cal S}[\omega]={\rm{Tr}} \bigl[G_{0}^{r} \Gamma_{L} G_{0}^{a} F \bigr].
\end{equation}
From the expression of ${\cal S}[\omega]$ it is clear that the average energy current due to driven force is independent of $\hbar$ and since it contains $G_{0}^{r,a}$ and $\Gamma_L$, which are independent of temperature we can conclude that the energy current is independent of the temperature of the heat baths in the ballistic transport case. However the second cumulant and similarly the higher ones do depend on temperature of the baths. The second cumulant can be written as
\begin{eqnarray}
\frac{\langle \langle  Q_{d}^{2} \rangle \rangle}{t_M} &=& \int \frac{d\omega}{4 \pi \hbar} \, (\hbar \omega)^{2} \, {\cal S}[\omega] \Bigl[(1+2\,f_{L})- \nonumber \\
&& 2 \,{\cal T}(\omega) (f_L-f_R) \Bigr].
\label{second-driven-cumulant}
\end{eqnarray}
Similarly all the higher cumulants can be obtained from the CGF and we can conclude that the distribution $P(Q_d)$ is not Gaussian.

\subsection{Classical limit}
In this section we will give the classical limit of the steady state expression for the CGF $\ln {\cal Z}^{s}(\xi)$ and $\ln {\cal Z}^{d}(\xi)$ given in Eq.~(\ref{eq-steady1}) and Eq.~(\ref{eq-driven-CGF}).

First of all we note that retarded and advanced Green's functions, i.e., $G_{0}^{r}$ and $G_{0}^{a}$ are similar both for quantum and classical case, so they stay the same when $\hbar \to 0$. We know that in the classical limit $f_{\alpha} \rightarrow  \frac{k_{B}T_{\alpha}}{\hbar \omega}$ and also $e^{ix}=1+ix + \frac{(ix)^{2}}{2} + \cdots$, where $x=\xi \hbar \omega$. Using this we obtain from Eq.~(\ref{eq-steady1}) the classical limit of ${\cal Z}^{s}(\xi)$. 
\begin{eqnarray}
\ln{\cal Z}^{s}_{\rm cls}(\xi)&=& \frac{t_M}{4\pi} \int d\omega\,\ln \det \Big[I-\big(G_{0}^{r} \Gamma_{L} G_{0}^{a} \Gamma_{R}\big) \times \nonumber \\
&& \, k_B T_L \, k_B T_R \,i\xi (i\xi +{\cal A})\Big].
\end{eqnarray}
This result reproduces that of Ref.~\cite{kundu}.
In the classical case also the CGF obeys the GC symmetry, i.e., it remains invariant under the transformation $i\xi \rightarrow -i\xi -{\cal A}$.

Let us now get the classical limit for $\ln {\cal Z}^{d}(\xi)$ using Eq.~(\ref{eq-driven-CGF}). Following above relations the function ${\cal K}(\xi)$ in the limit $\hbar \rightarrow 0$ reduces to
\begin{equation}
{\cal K}_{\rm cls}(\xi)= -\hbar \omega \,\Big(i\xi + \frac{\xi^2}{\beta_L}\Big).
\end{equation} 
The transmission function ${\cal S}[\omega]$ stays the same as it is independent of temperature and $\hbar$. So in the classical limit $\ln {\cal Z}^{d}(\xi)$ reduces to 
\begin{equation}
\ln {\cal Z}^{d}_{\rm cls}(\xi)= t_M \int \frac{d\omega}{4 \pi} \, \omega \, {\cal S}[\omega] \, \frac{\Big(i\xi + \frac{\xi^2}{\beta_L}\Big)}{{\cal N}_{\rm cls}(\xi)},
\end{equation}
where 
\begin{eqnarray}
{\cal N}(\xi)_{\rm cls}&=&\det\Big[I-\big(G_{0}^{r} \Gamma_{L} G_{0}^{a} \Gamma_{R}\big) \,k_B T_L \, k_B T_R \nonumber \\
&&i\xi (i\xi +{\cal A})\Big].
\end{eqnarray}
Here we can easily see that ${\cal Z}^{d}(-i\beta_L)=1$.

We can also derive the fluctuation dissipation theorem from Eq.(\ref{second-driven-cumulant}) if we assume the leads are at the same temperature, i.e., $\beta_{L}=\beta_{R}=\beta$ then we can write the second cumulant $\langle \langle Q_{d}^{2} \rangle \rangle$ as 
\begin{equation}
\frac{\langle \langle Q_{d}^{2} \rangle \rangle}{t_M} = \int \frac{d\omega}{4 \pi \hbar} \, (\hbar \omega)^{2} \,{\cal S}[\omega] (1+2\,f_{L}).
\end{equation}
In the high-temperature limit using $f_{L} \rightarrow \frac{k_{B}T_{L}}{\hbar \omega}$ and we obtain
\begin{equation}
\langle \langle Q_{d}^{2} \rangle \rangle =\frac{2}{\beta_{L}} \langle Q_{d} \rangle.
\end{equation}

In the next section we discuss Nazarov's generating function and give long-time limit expression. 

\section{Nazarov's definition of generating function}
In this section we will derive another definition of CGF given by Eq.~(\ref{eq-Z1-Nazarov}), starting from the CGF, derived using two-time measurement concept, i.e., Eq.~(\ref{eq-Z-two-time}). Eq.~(\ref{eq-Z1-Nazarov}) can be obtained from Eq.~(\ref{eq-Z-two-time}) in the small $\xi$ approximation as follows. In the small $\xi$ approximation the modified Hamiltonian given in Eq.~(\ref{modified}) takes the following form 
\be
{\cal H}_{x}(t)={\cal H}(t)+\hbar x {\cal I}_L(0),
\ee
because $\lim_{x\rightarrow 0}{\cal C}(x)=0$ and $\lim_{x\rightarrow 0}{\cal S}(x)=\hbar x V^{LC}$. ${\cal I}_L$ is defined in Eq.~(\ref{current}).
So the modified unitary operator becomes
\be
{\cal U}_{x}(t,0)=T e^{-\frac{i}{\hbar} \int_{0}^t [{\cal H}(\bar{t})+\hbar x {\cal I}_L(0)] d\bar{t}}.
\ee
We can consider $\hbar x {\cal I}_L(0)$ as the interaction Hamiltonian and write the full unitary operator ${\cal U}_x$ as a product of two unitary operators as following 
\begin{equation}
{\cal U}_{x}(t,0)={\cal U}(t,0) \, {\cal U}_{x}^{I}(t,0),
\label{u}
\end{equation}
where 
\begin{eqnarray}
{\cal U}(t,0)&=& T e^{-\frac{i}{\hbar} \int_{0}^t {\cal H}(t') \, dt'}, \nonumber \\
{\cal U}_{x}^{I}(t,0) &=& T e^{-\frac{i}{\hbar} \int_{0}^t  \hbar x {\cal I}_L(t') dt'},
\label{u1}
\end{eqnarray}
with ${\cal I}_L(t')={\cal U}^{\dagger}(t',0)\,{\cal I}_L(0)\,{\cal U}(t',0)$ is the current operator in the Heisenberg picture. It is important to note that ${\cal U}$ is the usual unitary operator which evolves with the full Hamiltonian ${\cal H}(t)$ in Eq.~(\ref{eq-unitary}) and has no $\xi$ dependence.

If we use product state $\rho(-\infty)$ as the initial state the CGF is given by 
\begin{equation}
{\cal Z}(\xi)={\rm Tr}\big[\rho(-\infty) \, {\cal U}_{\xi/2}(0,t)\, {\cal U}_{-\xi/2}(t,0)\big].
\end{equation} 
In the small $\xi$ approximation and using the expressions for ${\cal U}_{x}$ we can write the CGF as
\begin{equation}
{\cal Z}_{1}(\xi)=\lim_{\xi \to 0} {\cal Z}(\xi)= {\rm Tr}\big[\rho(-\infty) \, {\cal U}_{\xi/2}^{I}(0,t)\, {\cal U}_{-\xi/2}^{I}(t,0)\big],
\end{equation}
where we use the property of unitary operator, i.e., ${\cal U}^{\dagger}(t,0) {\cal U}(t,0)=1$. Finally using the definition of heat operator ${\cal Q}_L$ given in Eq.~(\ref{eq-hatQ}) and the CGF can be written down as 
\begin{equation}
{\cal Z}_{1}(\xi)=\Big \langle {\bar T} e^{i\xi {\cal Q}_{L}(t)/2} \, T e^{i\xi {\cal Q}_{L}(t)/2}\Big \rangle,
\end{equation}
which is the same as in Eq.~(\ref{eq-Z1-Nazarov}).

In the following we will give the long-time limit expression for this CGF.
 
In order to calculate the CGF, it is important to go to the interaction picture with respect to the Hamiltonian ${\cal H}_{0}={\cal H}_L+{\cal H}_C+ {\cal H}_R$, as we know how to calculate Green's functions for operators which evolves with ${\cal H}_{0}$ and treat the rest part as the interaction ${\cal V}_{x}={\cal H}_{\rm int}+ \hbar x {\cal I}_{L}(0)$.  So the CGF on contour $C=\big[0,t_M \big]$ can be written as
\begin{equation}
{\cal Z}_{1}(\xi)=\Big \langle T_c e^{-\frac{i}{\hbar} \int {\cal V}_{x}^{I}(\tau) d\tau } \Big \rangle,
\end{equation}
where ${\cal V}_{x}^{I}(\tau)$ is now given by 
\begin{eqnarray}
{\cal V}_{x}^{I}(\tau)&=&u_{L}^T(\tau) V^{LC} u_{C}(\tau)+ u_{R}^T(\tau) V^{RC} u_{C}(\tau) \nonumber \\
&&+\hbar x(\tau) p_{L}(\tau) V^{LC} u_{C}(\tau),
\end{eqnarray} 
where $p_L=\dot{u}_L$. The time-dependence $\tau$ is coming from the free evolution with respect to ${\cal H}_{0}$. $x(\tau)$ has the similar meaning as before, i.e., on the upper branch of the contour $x^{+}(t)=-\xi/2$ and on the lower branch $x^{-}(t)=\xi/2$. Now using the same idea as before, we expand the series, use Wick's theorem and finally the CGF can be expressed as
\begin{equation}
\ln {\cal Z}(\xi)=- \frac{1}{2} {\rm Tr}_{j,\tau} \ln \Big[ 1 - G_0 \Sigma_L^{A} \Big]. 
\end{equation}
Here $G_{0}$ is the same as before and is given by Eq.~(\ref{eq-Dyson-product}). However the shifted self-energy $\Sigma_{L}^{A}$ in this case is different and is given by (in contour-time argument) 
\begin{eqnarray}
\Sigma_{L}^{A}(\tau,\tau')&=&\hbar\, x(\tau)\, \Sigma_{p_L u_L}(\tau,\tau')+\hbar\, x(\tau')\, \Sigma_{u_L p_L}(\tau,\tau') \nonumber \\
&&+\hbar^{2}\, x(\tau)\,x(\tau')\,\Sigma_{p_L p_L}(\tau,\tau').
\end{eqnarray}
The notation $\Sigma_{A B}(\tau,\tau')$ means
\begin{equation}
\Sigma_{A B}(\tau,\tau')= \bigl(-\frac{i}{\hbar}\bigr) V^{CL} \,  \langle\, T_{c} A(\tau) B^{T}(\tau')\,\rangle \, V^{LC}.
\end{equation}
The average here is with respect to equilibrium distribution of the left lead. It is possible to express the correlation functions such as $\Sigma_{p_L u_L}(\tau,\tau')$ in terms of the $\Sigma_{u_L,u_L}(\tau,\tau')=\Sigma_L(\tau,\tau')$ correlations. $\Sigma_{p_L u_L}(\tau,\tau')$ and $\Sigma_{u_L p_L}(\tau,\tau')$ is simply related with $\Sigma_L(\tau,\tau')$ by the contour-time derivative whereas for  $\Sigma_{p_L p_L}(\tau,\tau')$ the expression is 
\begin{equation}
\Sigma_{p_L p_L}(\tau,\tau')= \frac{\partial^{2} \Sigma_{u_L u_L}(\tau,\tau')}{\partial \tau \partial \tau'} + \delta(\tau,\tau') \Sigma_{L}^{I}.
\end{equation}
Where $\Sigma_L^{I}=V^{CL} V^{LC}$. Now in the frequency domain different components of $\Sigma_L^{A}$ takes the following form
\begin{eqnarray}
\Sigma_A^{t}[\omega]&=& \frac{\hbar^{2}\xi^{2}\omega^{2}}{4} \Sigma_{L}^{t}[\omega]+ \frac{\hbar^{2}\xi^{2}}{4}\Sigma_{L}^{I}, \nonumber \\
\Sigma_A^{\bar{t}}[\omega]&=& \frac{\hbar^{2}\xi^{2}\omega^{2}}{4} \Sigma_{L}^{\bar{t}}[\omega]- \frac{\hbar^{2}\xi^{2}}{4}\Sigma_{L}^{I}, \nonumber \\
\Sigma_A^{<}[\omega]&=& \big( i \hbar \xi \omega - \frac{\hbar^{2}\xi^{2}\omega^{2}}{4} \big) \Sigma_{L}^{<}[\omega], \nonumber \\
\Sigma_A^{>}[\omega]&=& \big(-i \hbar \xi \omega - \frac{\hbar^{2}\xi^{2}\omega^{2}}{4} \big) \Sigma_{L}^{>}[\omega]. 
\end{eqnarray}

Finally using the relations between the self-energy (see Appendix A), in the long-time limit the CGF can be written down as,
\begin{eqnarray}
\ln {\cal Z}_{1}(\xi) &=& - t_M \int \frac{d\omega}{4\pi} \ln \Big[1-(i\xi \hbar \omega){\cal T}[\omega]\,(f_L-f_R) 
\nonumber \\ &&- \frac{(i\xi \hbar \omega)^{2}}{4}\Big({\cal T}[\omega](1+2f_L)(1+2f_R)-G_{0}^{a}\Sigma_{L}^{r} \nonumber \\
&&+G_{0}^{r}\Sigma_{L}^{a} -G_{0}^{r}\Gamma_{L}G_{0}^{a}\Gamma_{L}\Big)+ {\cal J}(\xi^{2},\xi^{4})\Big],
\end{eqnarray}
where ${\cal J}(\xi^2,\xi^4)$ is given by
\begin{eqnarray}
{\cal J}(\xi^2,\xi^4)&=&-\frac{\hbar^2 \xi^2}{4}\big(G_{0}^{a}+G_{0}^{r}\big)\Sigma_{L}^{I} - \frac{1}{4} \frac{(i\xi \hbar \omega)^2}{2} \frac{\hbar^2 \xi^2}{2} \nonumber \\ 
&&+\big(G_{0}^{r}\Sigma_{L}^{a}G_{0}^{a}\Sigma_L^{I}+G_{0}^{r}\Sigma_{L}^{I}G_{0}^{a}\Sigma_L^{r}\big) + \frac{1}{4} \frac{(i\xi \hbar \omega)^4}{4} \nonumber \\
&&G_{0}^{r}\Sigma_{L}^{a}G_{0}^{a}\Sigma_L^{r} + \frac{1}{4} \frac{(\hbar^{4} \xi^{4})}{4} G_{0}^{r}\Sigma_{L}^{I}G_{0}^{a}\Sigma_L^{I}.
\end{eqnarray}

This CGF does not obey the GC fluctuation symmetry. However it gives the correct first and second cumulant as it should because the definition of first and second cumulant turn out to be the same for both the generating functions ${\cal Z}(\xi)$ and ${\cal Z}_{1}(\xi)$ and is given by
\begin{eqnarray}
&&\langle \langle Q \rangle \rangle=\langle Q \rangle = \frac{\partial \ln {\cal Z}(\xi)}{\partial {(i\xi)}}=\frac{\partial \ln {\cal Z}_1(\xi)}{\partial {(i\xi)}}= \int_{0}^{t}  dt_1 \langle {\cal I}_L(t_1) \rangle, \nonumber \\
&&\langle \langle Q^{2} \rangle \rangle =\langle Q^{2} \rangle- \langle Q \rangle^{2} = \frac{\partial^{2} \ln {\cal Z}(\xi)}{\partial {(i\xi)^{2}}}=\frac{\partial^{2} \ln {\cal Z}_1(\xi)}{\partial {(i\xi)^{2}}}\nonumber \\
&& \> = \int_{0}^{t} dt_1  \int_{0}^{t} dt_2  \langle {\cal I}_L(t_1) {\cal I}_L(t_2) \rangle-\Big[\int_{0}^{t}\! dt_1\! \langle {\cal I}_L(t_1) \rangle \Big]^{2}. 
\end{eqnarray}
Expressions for higher cumulants are different for the two generating functions and hence the final expressions for the CGF's are completely different from each other.  

\section{Conclusion}
In summary, we present an elegant way of deriving the CGF for heat ${\cal Q}_{L,R}$  transferred from the leads to the center for driven linear systems using the two-time measurement concept and with the help of the NEGF technique. The CGF is written in terms of the Green's function of the center and the  self-energy $\Sigma_{L}^A$ of the leads. 
The counting of the energy is related to the shifting in time for the self-energy.
This expression is valid in both transient and steady state regimes, where the information about the measurement time $t_M$ is contained in $\Sigma_{L}^A$.  The form of the expression,
$-(1/2) {\rm Tr} \ln (1 - G_{0} \Sigma_L^A)$, is the same whether we use a product initial state or a projected initial state, except that the meaning of the Green's function has to be adjusted accordingly. We consider three different initial conditions and show numerically for 1D linear chains connected with Rubin baths, that transient behaviors significantly differs from each other but eventually leads to the same steady state distribution in the long-time limit. We give explicit expressions of the CGF in the steady state invoking the symmetry of translational invariance in time. The CGF obeys the GC symmetry. We also give the steady state expression for the CGF in the presence of time-dependent driving forces. We obtain a two parameter CGF which is useful for calculating the correlations between heat flux and also the total entropy which flows to the leads. Our calculations can be easily generalized to arbitrary dimensions with any number of heat baths. We will show in the appendix that our method can be extended for the electronic calculations where we derive the CGF for a tight-binding model. It will be interesting to derive the CGF by taking magnetic field contribution into the Hamiltonian and also to study the cumulants in the presence of nonlinear interactions such as phonon-phonon interactions or electron phonon interactions.

\section*{Acknowledgments}
We are grateful to Juzar Thingna, Meng Lee Leek, Zhang Lifa, and Li Huanan for insightful discussions. This work is supported in part by a URC research grant R-144-000-257-112 of National University of Singapore.

\section*{Appendix}
\subsection{Expressions for different type of Green's functions}
Here we give the explicit expressions for the center Green's function $G_{0}[\omega]$ in the steady state, for a harmonic system which is connected with the leads. These formulas are required to derive the analytical form of the CGF given in Eq.~(\ref{eq-steady1}). For the basic definitions of different types of Green's functions we refer to Ref.~\onlinecite{WangJS-europhysJb-2008}. 

The retarded Green's function $G_{0}^{r}[\omega]$ is given by
\begin{equation}
G_{0}^{r}[\omega]=\Big[(\omega+i\eta)^{2}-K^{C}-\Sigma_{L}^{r}[\omega]-\Sigma_{R}^{r}[\omega]\Big]^{-1}.
\end{equation}
Here $\eta$ is an infinitesimal positive number which is required to satisfy the condition of causality i.e.,$G_{0}^r(t)=0$ for $t <0 $.
The advanced Green's function is $G_{0}^{a}[\omega]=\big[G_{0}^{r}[\omega]\big]^{\dagger}$. The Keldysh Green's function $G_{0}^{K}[\omega]$ can be obtained by solving the corresponding Dyson equation, Eq.~(\ref{eq-Dyson-product}), and is given by
\begin{equation}
G_{0}^{K}[\omega]=G_{0}^{r}[\omega]\Sigma^{K}[\omega]G_{0}^{a}[\omega],
\end{equation}
where $\Sigma^{K}=\Sigma_L^{K}+\Sigma_R^{K}$ and $\Sigma_{\alpha}^{K}=\Sigma_{\alpha}^{<}+ \Sigma_{\alpha}^{>}$ with $\alpha=L,R$. Alternatively, $G_{0}^{K}=G_{0}^< +G_{0}^>$. Another important identity is
\begin{equation}
G_{0}^{r}[\omega]-G_{0}^{a}[\omega]=-i\, G_{0}^{r}[\omega] \big(\Gamma_L[\omega]+\Gamma_R[\omega]\big)G_{0}^{a}[\omega],
\end{equation}
where $\Gamma_{\alpha}[\omega]=i\big(\Sigma_{\alpha}^{r}[\omega]-\Sigma_{\alpha}^{a}[\omega]\big)$, and $\alpha=L,R$. The self-energy for the leads are given by
\begin{eqnarray}
\Sigma_{\alpha}^{<}[\omega]&=&f_{\alpha}[\omega]\big(\Sigma_{\alpha}^{r}[\omega]-\Sigma_{\alpha}^{a}[\omega]\big), \nonumber \\
\Sigma_{\alpha}^{>}[\omega]&=&(1+f_{\alpha}[\omega])\big(\Sigma_{\alpha}^{r}[\omega]-\Sigma_{\alpha}^{a}[\omega]\big).
\end{eqnarray}
where $f_{\alpha}[\omega]=1/\bigl(e^{\beta_{\alpha} \hbar \omega_{\alpha}}-1\bigr)$ is the Bose distribution function. 

Explicit expressions for $G_{0}^{r}[\omega]$ and $\Sigma_{L}^{r}[\omega]$ can be obtained for 1D homogeneous linear chain, with inter particle force constant $K$ and onsite spring constant $K_{0}$ and which is divided into three parts: the center, the left and the right. The classical equation of motion for the atoms in all three regions is
\begin{equation}
\ddot{u}_j=K u_{j-1} + \bigl (-2K -K_{0}\bigr )u_{j} + Ku_{j-1},
\end{equation}
where the index $j$ runs over all the atoms in the full system.

The retarded Green's function $G_{0}^{r}[\omega]$ can be obtained by solving \cite{Wang-pre07} $[(\omega+i\eta)^{2}-\tilde{K}]G_{0}^{r}=I$, where matrix $\tilde{K}$ which is infinite in both directions and is $2K+K_{0}$ on the diagonals and $-K$ on the first off-diagonals.  The solution is translationally invariant in space index and is given by 
\be
G_{0,jk}^{r}[\omega]=\frac{\lambda^{|j-k|}}{K(\lambda-\frac{1}{\lambda})},
\ee 
with $\lambda=-\frac{\Omega}{2K}\pm \frac{1}{2K}\sqrt{\Omega^{2}-4K^{2}}$ and $\Omega=(\omega+i\eta)^{2}-2K-K_{0}$, choosing between plus and minus sign by $|\lambda|\le 1$. 

The surface Green's function $g_L^{r}[\omega]$ can be similarly obtained in frequency domain and is given in terms of the self-energy $\Sigma_{L}^{r}[\omega]=-K \lambda$. Since in equilibrium only one Green's function is independent, knowing $\Sigma_{L}^{r}[\omega]$ is sufficient to obtain all other Green's functions.

Here we also give the expressions for Green's functions $g_C$ in time and frequency domain for an isolated single harmonic oscillator with frequency $\omega_{0}$ (we have omitted the subscript $C$ in $g_C$) \cite{Zeng,Brouwer}
\begin{eqnarray}
g^{r}(t) & = & -\theta(t) \, \frac{\sin{\omega_{0}t}}{\omega_{0}},\nonumber \\
g^{r}[\omega] & = & \frac{1}{(\omega+i\eta)^{2}-\omega_{0}^{2}},\nonumber \\
g^{<}(t) & = & \frac{-i}{2\omega_{0}}\left[(1+f)e^{i\omega_{0}t}+fe^{-i\omega_{0}t}\right],\nonumber \\
g^{<}[\omega] & = & \frac{-i\pi}{\omega_{0}}\left[\delta(\omega+\omega_{0})(1+f)+\delta(\omega-\omega_{0})f\right],
\end{eqnarray} 
where $f=f(\omega_{0})=\frac{1}{e^{\beta\hbar\omega_{0}}-1}$. Other components can be obtained by exploiting the symmetry between the Green's functions such as $g^a(-t)=g^r(t)$ for $t > 0$ hence $g^r[\omega]=g^a[-\omega]$. The greater component is related with the lesser component via $g^>(t)=g^<(-t)$ which in the frequency domain satisfy $g^>[\omega]=g^<[-\omega]$.



\subsection{Current at short time for product initial state $\rho(-\infty)$}
Using the definition of current operator given in Eq.~(\ref{current}) the energy current flowing from the left lead to the center is (here we assume that there is no driving force $f(t)$) 
\begin{equation}
\langle {\cal I}_{L}(t)\rangle =-\langle \frac{d{\cal H}_L(t)}{dt} \rangle =\frac{i}{\hbar} \langle \big[{\cal H}_{L}(t),{\cal H} \big] \rangle ,
\end{equation}
where the average is with respect to $\rho(-\infty)$. If $t$ is small we can expand ${\cal H}_{L}(t)$ in a Taylor series and is given by ${\cal H}_{L}(t)={\cal H}_{L}(0)+t \dot{{\cal H}}_{L}(0) +\cdots $ 

Now since $\big[\rho(-\infty), {\cal H}_{L}(0) \big]=0$, then it immediately follows that $\langle \big[{\cal H}_{L}(0),{\cal H}\big]\rangle =0$ by using the cyclic property of trace. So in linear order of $t$ the current is given by
\begin{equation}
\langle {\cal I}_{L}(t)\rangle=t \frac{i}{\hbar} \langle \big[{\dot{\cal H}}_{L}(0),{\cal H}\big]\rangle= -t \frac{i}{\hbar}\langle \big[p_{L}^{T}V^{LC}u_C, {\cal H}\big]\rangle.
\end{equation}
The only term of full ${\cal H}$ that will contribute to the  is ${\cal H}_{LC}=u_{L}^T V^{LC} u_{C}$. 

Now using the relation that $\big[p_L,u_L\big]=-i\hbar$, for one-dimensional linear chain we can write
\begin{equation}
\langle {\cal I}_{L}(t) \rangle =-t \, K^{2} \langle (u^{C}_{1})^{2} \rangle = -t \, K^{2} \frac{\hbar}{\omega_{0}} \Big(f_{C}(\omega_{0})+\frac{1}{2}\Big).
\end{equation}
where $u^{C}_{1}$ is the first particle in the center which is connected with the first particle of the left lead with force constant $K$. Now since the average is with respect to $\rho(-\infty)$,$\langle (u^{C}_{1})^{2} \rangle $ can be easily computed. Here $f_{C}(\omega_{0})$ is the Bose distribution function of the particle with characteristic frequency $\omega_{0}$. So we can see that for short time the current is negative, i.e, it goes into the lead. It is now easy to see that similar expression should also hold for $\langle {\cal I}_{R}(t)\rangle$.  The negative sign in currents means that the energy flows into the leads initially irrespect to the temperature of the leads. This is consistent with the numerical results obtained by Cuansing et al.\ \cite{eduardos-paper,eduardos-paper1}.

\subsection{\label{apdC} Convolution, trace, and determinant on Keldysh contour}
Here we discuss the meaning of convolution, trace and determinant on the Keldysh contour which we used to derive the CGF's for heat flux. We define the convolution on contour in the following way.
\begin{eqnarray}
A B  \cdots D &\rightarrow& \sum_{j_2,j_3, \cdots, j_n}\int d\tau_2 \cdots \int d\tau_n A_{j_1,j_2}(\tau_1, \tau_2)\nonumber \\
&& B_{j_2,j_3}(\tau_2, \tau_3) \cdots D_{j_n,j_{n+1}}(\tau_n, \tau_{n+1}), 
\end{eqnarray}
From the convolution we define trace by substituting $\tau_{n+1}=\tau_1$, $j_{n+1}=j_1$ and integrate also over $\tau_1$, sum over $j_1$ i.e., 
\begin{eqnarray}
{\rm Tr}_{j,\tau}(AB \cdots D)&=&\int d\tau_1 \int d\tau_2 \cdots \int d\tau_n \\
&& {\rm Tr}_j\bigl[ A(\tau_1, \tau_2) B(\tau_2, \tau_3) \cdots D(\tau_n, \tau_1) \bigr], \nonumber  
\end{eqnarray}
Changing from contour to real-time integration from $-\infty$ to $+\infty$, i.e., using $\int d\tau = \sum_{\sigma} \int \sigma dt $ we have
\begin{eqnarray}
{\rm Tr}_{j,\tau}(AB \cdots D) = \!\!\!\sum_{\sigma_1,\sigma_2, \cdots, \sigma_n}\!\!\! \int dt_1 \int dt_2 \cdots \int dt_n \qquad  \\
{\rm Tr}_j \bigl[ A^{\sigma_1\sigma_2}(t_1, t_2) \sigma_2 B^{\sigma_2\sigma_3}(t_2, t_3) \cdots \sigma_n D^{\sigma_n\sigma_{n+1}}(t_n, t_{1}) \bigr]. \nonumber   
\end{eqnarray}
Let us absorb the extra $\sigma$ into the definition of branch components,
i.e., define
\begin{equation}
\bar{A}_{\sigma\sigma'} = \sigma A^{\sigma\sigma'}, \quad{\rm or}\quad
\bar{A} = \sigma_z A,
\end{equation}
where $A$ is viewed as $2\times 2$ block matrix with the usual
$+$, $-$ component, 
\begin{equation}
A = \left( \begin{array}{cc}
                             A^{++} & A^{+-} \\
                             A^{-+} & A^{--} 
                         \end{array} \right) = 
\left( \begin{array}{cc}
                             A^t & A^< \\
                             A^> & A^{\bar t} 
                         \end{array} \right),
\end{equation}
and $\sigma_z$ is defined as
\begin{eqnarray}
\sigma_z &=& \left( \begin{array}{cc}
                             1 & 0 \\
                             0 & -1 
                         \end{array} \right),
\end{eqnarray}
then it can be easily seen that 
\begin{eqnarray}
{\rm Tr}_{j,\tau}(AB \cdots D)&=& \int dt_1 \int dt_2 \cdots \int dt_n  {\rm Tr}_{j} \bigl[{\bar A}(t_1,t_2) \nonumber \\
&&{\bar B}(t_2,t_3) \cdots {\bar D}(t_n,t_1) \bigr] \nonumber \\
&&={\rm Tr}_{t,j,\sigma}(\bar{A}\bar{B} \cdots \bar {D}).
\end{eqnarray}
Then we can do a rotation, where the rotation matrix is given by 
\begin{eqnarray}
O &=& \frac{1}{\sqrt{2}} \left(
\begin{array}{cc}
                             1 & 1 \\
                            -1 & 1 
\end{array} \right), \quad OO^T = I.
\end{eqnarray}
and we define for any matrix $A$, the rotated matrix as 
\begin{equation}
\breve{A} = O^T \sigma_z A O = O^{T} \bar{A} O.
\end{equation}
This is known as Keldysh rotation. The effect of Keldysh rotation is given in Eq.~(\ref{keldysh-rotation}). Since this is an orthogonal transformation the trace remains invariant and hence we can write
\begin{eqnarray}
{\rm Tr}_{t,j,\sigma}(\bar{A}\bar{B} \cdots \bar {D})&=& {\rm Tr}_{t,j,\sigma}(\breve{A}\breve{B} \cdots \breve{D}).
\end{eqnarray}
If we now go to the frequency domain using the definition of two-time Fourier transform given in Eq.~(\ref{two-time-FT}) 
then we can compute the trace in frequency domain as 
\begin{eqnarray}
{\rm Tr}_{(j,\tau)}(AB \cdots D) &=& \int\! \frac{d\omega_1}{2\pi} \!
\int\! \frac{d\omega_2}{2\pi}\! \cdots
\int\! \frac{d\omega_n}{2\pi} \!
{\rm Tr} \bigl\{ \nonumber \\
 && \breve{A}[\omega_1, -\omega_2] 
\breve{B}[\omega_2, -\omega_3] \cdots \breve{D}[\omega_n, -\omega_1] \bigr\}
\nonumber \\ 
&=& {\rm Tr}_{j,\sigma,\omega} ( \breve{A} \breve{B} \cdots \breve{D} ).
\end{eqnarray}
The last line above define what we mean by trace over frequency domain given in Eq.~(\ref{eq-Zsteady}).
Unlike trace in time domain, the second argument of the each of the variables
need a minus sign. 

Let us now define what do we mean by 1 on contour. In the sense of convolution we define 1 as
\begin{equation}
A \, 1\, D =A \, D
\end{equation}
which means
\begin{equation}
\int d\tau_1 \!\! \int d\tau_2\, A(\tau,\tau_1) I \delta(\tau_1,\tau_2) D(\tau_2,\tau')= \int d\tau_1 A(\tau,\tau_1) D(\tau_1,\tau'). 
\end{equation}
Note that $\delta(\tau,\tau')$ in the real time has the following form
\begin{equation}
\delta^{\sigma,\sigma'}(t,t')=\sigma \delta_{\sigma,\sigma'} \delta(t-t').
\end{equation}
The inverse on the contour is defined as
\begin{equation}
\int d\tau_1 A(\tau,\tau_1) B(\tau_1,\tau') = I \delta(\tau,\tau'),
\end{equation}
where the identity matrix $I$ takes care about the space index. Similar to the above we go to the real time and multiply the above equation with the branch index $\sigma$ and we can write,
\begin{equation} 
\int dt_1 {\bar A}(t,t_1) {\bar B}(t_1,t') = I \bar{\delta}(t-t').
\end{equation}
where 
\begin{eqnarray}
\bar{\delta}(t-t')&=&\sigma \delta^{\sigma,\sigma'}(t,t')=\sigma^{2} \delta_{\sigma,\sigma'} \delta(t-t') \nonumber \\
&&=\delta_{\sigma,\sigma'} \delta(t-t')
\end{eqnarray}
If we now discretize the time and write $\delta(t_i,t_{i'})=\delta_{i,i'}/\Delta t$ with $\Delta t= |t_i-t_{i'}|$ then we have
\begin{equation}
\tilde{A} \tilde{B} =\tilde{I}.
\end{equation}
with $\tilde{A}= A \Delta t$ and similarly for other matrices. 

With similar notions we can now write different types of Dyson's equation given in Eq.~(\ref{eq-Dyson-full},\ref{eq-Dyson-product}) as following. In contour time we have 

\begin{eqnarray}
G_0(\tau, \tau') &=& g_C(\tau,\tau') \\
&&\> + \int \!\int d\tau_1d \tau_2\, 
g_C(\tau, \tau_1) \Sigma(\tau_1,  \tau_2) G_0(\tau_2, \tau'), \nonumber
\end{eqnarray}
In real time following the above arguments we write
\begin{eqnarray}
\bar{G}_{0}(t,t') &=& \bar{g}_C(t,t') \\
&&\> + \int \!\int dt_1 dt_2\, 
{\bar g}_C(t, t_1) {\bar \Sigma}(t_1, t_2) {\bar G}_0(t_2, t'), \nonumber
\end{eqnarray}
After Keldysh rotation we can write 
\begin{eqnarray}
\breve{G}_{0}(t,t') &=& \breve{g}_C(t,t') \\
&&\> + \int \!\int dt_1 dt_2\, 
{\breve g}_C(t, t_1) {\breve \Sigma}(t_1, t_2) {\breve G}_0(t_2, t'). \nonumber
\end{eqnarray}
Finally in the discretize time $t$ we write 
\begin{equation}
\tilde{G}_{0}=\tilde{g}_C + \tilde{g}_C \tilde{\Sigma} \tilde{G}_{0},
\end{equation}
which is a matrix equation. Similar equations can also be written down for Eq.~(\ref{eq-Dyson-full}).
 
Now we define determinant via the relation $\det(A)=\exp({\rm Tr} \ln A)$, i.e, the determinant is defined in terms of trace. In order for $\ln A$ to be defined we have to assume a Taylor expansion. For example we can define $\ln(1+M)=M-M^2/2 + M^3/3 + \cdots $ where 1 means $\delta_{jj'}\delta(\tau,\tau')$ in contour space.

\subsection{A quick derivation of the Levitov-Lesovik formula for electrons using NEGF}
The generating function for the non-interacting electrons was first derived by Levitov and Lesovik \cite{Levitov,Levitov1} using Landauer type of wave scattering approach.  Klich \cite{Klich} and Sch\"onhammer \cite{otherworks2} re-derived the formula using a trace and determinant relation to reduce the problem from many-body problem to a single particle Hilbert space problem. Esposito et al.\ gave an approach using the superoperator nonequilibrium Green's function \cite{Esposito-review-2009}. A more rigorous treatment is given in Ref.~\onlinecite{Bernard}.

Our method for calculating CGF can be easily extended for the electron case. Here we will derive the CGF for the joint probability distribution for particle and energy without time-dependent driving force. 
The Hamiltonian of the whole system can be written as (using tight-binding model)
\begin{equation}
{\cal H}^{e}=\sum_{\alpha=L,C,R} c_{\alpha}^{\dagger} h^{\alpha} c_{\alpha} + \sum_{\alpha=L,R} \big(c_{\alpha}^{\dagger} V_{e}^{\alpha C} c_{C} + {\rm h.c.}\big)
\end{equation}
where $c_{\alpha}$ is a column vector consisting of all the annihilation operator of region $\alpha$. $c_{\alpha}^{\dagger}$ is a row vector of  the corresponding creating operators. $h^{\alpha}$ is the single particle Hamiltonian matrix. $V_{e}^{\alpha C}$ has similar meaning as $V^{\alpha C}$ in the phonon Hamiltonian and $V_{e}^{\alpha C}=(V_{e}^{C\alpha})^{\dagger}$. 

We are interested in calculating the generating function corresponding to the particle operator ${\cal N}_L$ and energy operator ${\cal H}_L$ of the left-lead where ${\cal H}_L= c_{L}^{\dagger} h^{L} c_L$ and ${\cal N}_L= c_{L}^{\dagger} c_L$ \cite{Hanggi1}. One can easily generalize the formula for right lead also as we did in the phonon case. For electrons ${\cal N}_L$ and ${\cal H}_L$ can be measured simultaneously because they commute, i.e., $\big[{\cal H}_L, {\cal N}_L \big]=0$. In order to calculate the CGF we introduce two counting fields $\xi_p$ and $\xi_e$ for particle and energy respectively. Here we will consider the product initial state (with fixed temperatures and chemical potentials for the leads) and derive the long-time result.

Similar to the phonon case we can write the CGF as 
\begin{equation}
{\cal Z}(\xi_e,\xi_p)= \Big \langle e^{i\big(\xi_e {\cal H}_L + \xi_p {\cal N}_L \big)}\,e^{-i\big(\xi_e {\cal H}^{H}_L + \xi_p {\cal N}^{H}_L \big)} \Big \rangle,
\end{equation}
where superscript $H$ means the operators are in the Heisenberg picture at time $t$. In terms of modified Hamiltonian the CGF can be expressed as
\begin{equation}
{\cal Z}(\xi_e,\xi_p)= \Big \langle {\cal U}_{(\frac{\xi_e}{2},\frac{\xi_p}{2})} (0,t) \, {\cal U}_{(-\frac{\xi_e}{2},-\frac{\xi_p}{2})} (t,0) \Big \rangle,
\end{equation} 
where 
\begin{eqnarray}
{\cal U}_{x,y}(t,0)&=& e^{i x {\cal H}_L + i y {\cal N}_L}\, {\cal U}(t,0)\,e^{-i x {\cal H}_L - i y {\cal N}_L} \nonumber \\
&&= e^{-\frac{i}{\hbar} {\cal H}_{x,y} t}
\end{eqnarray}
with $x=\xi_e/2$ and $y=\xi_p/2$ and ${\cal U}(t,0)=e^{-i {\cal H} t}$. ${\cal H}_{x,y}$ is the modified Hamiltonian which evolves with both ${\cal H}_L$ and ${\cal N}_L$ and is given by
\begin{eqnarray}
{\cal H}_{x,y}&=& e^{i x {\cal H}_L + i y {\cal N}_L}\, {\cal H} \,e^{-i x {\cal H}_L - i y {\cal N}_L} \nonumber \\
&&= {\cal H}_L + {\cal H}_C + {\cal H}_R + \big(e^{iy} c_{L}^{\dagger}(\hbar x) V_{e}^{LC} c_C +{\rm h.c.}\big)\nonumber \\
&& + \big(c_R^{\dagger} V_{e}^{RC} c_{C} + {\rm h.c.}\big),
\end{eqnarray}
where we have used the fact that
\begin{eqnarray}
e^{i x {\cal H}_L} c_{L}(0) e^{-i x {\cal H}_L} &=& c_{L}(\hbar x), \nonumber \\
e^{i y {\cal N}_L} c_{L}(0) e^{-i y {\cal N}_L} &=& e^{-i y} c_{L}.
\end{eqnarray}
So the evolution with ${\cal H}_L$ and ${\cal N}_L$ is to shift the time-argument and produce a phase for $c_{L},c_{L}^{\dagger}$ respectively. Next we go to the interaction picture of the modified Hamiltonian ${\cal H}_{x,y}$ with respect to ${\cal H}_{0}= \sum_{\alpha=L,C,R} {\cal H}_{\alpha}$ and the CGF then can be written on the contour running from 0 to $t_M$ and back as,
\begin{equation}
{\cal Z}(\xi_e,\xi_p)= {\rm Tr}\Big[\rho(-\infty) T_{c} e^{-\frac{i}{\hbar} \int d\tau {\cal V}_{x,y}^{I}(\tau)} \Big],
\end{equation} 
where ${\cal V}_{x,y}^{I}(\tau)$ is written in contour time. 
\begin{eqnarray}
{\cal V}_{x,y}^{I}(\tau)&=&\big ( e^{i y} c_{L}^{\dagger}(\tau+\hbar x) V_{e}^{LC} c_C (\tau)+ {\rm h.c.} \big)+ \nonumber \\
&& \big(c_{R}(\tau)^{\dagger} V_{e}^{RC} c_C(\tau) + {\rm h.c.}\big).
\end{eqnarray}
Now we can expand the exponential in the generating function and use Feynman diagrams to sum the series and finally the CGF can be shown to be
\begin{equation}
\ln {\cal Z}(\xi_e,\xi_p)={\rm Tr}_{j,\tau} \ln \Big[1-G^{e}_{0} \Sigma_{L,e}^{A} \Big],
\end{equation}
where we define the shifted self-energy for the electron case as
\begin{equation}
\Sigma_{L,e}^{A}(\tau,\tau')=e^{i (y(\tau')-y(\tau))} \Sigma_{L,e}(\tau+\hbar x, \tau'+\hbar x') -\Sigma_{L,e}(\tau,\tau').
\end{equation}
The counting of the electron number is associated with factor of a phase, while
the counting of the energy is related to translation in time.
Note that the CGF does not have the characteristic $1/2$ pre-factor as compared to the phonon case because $c$ and $c^{\dagger}$ are independent
variables. In the long-time limit following the same steps as we did for phonons, the CGF can be written down as (after doing Keldysh rotation) 
\begin{equation}
\ln {\cal Z}(\xi_e,\xi_p)=t_M \int \frac{dE}{2\pi \hbar} {\rm Tr} \ln \big(I- \breve{G}^{e}_{0}(E)\breve{\Sigma}_{L,e}^{A}(E)\big).
\end{equation}
In the energy $E$ domain different components of the shifted self-energy are
\begin{eqnarray}
\Sigma_{A}^{t}(E)&=&\Sigma_{A}^{\bar{t}}(E)=0, \nonumber \\
\Sigma_{A}^{<}(E)&=& \big( e^{i (\xi_p + \xi_e E)} -1 \big) \Sigma_{L}^{<}(E), \nonumber \\
\Sigma_{A}^{>}(E)&=& \big( e^{-i (\xi_p + \xi_e E)} -1 \big) \Sigma_{L}^{>}(E). 
\end{eqnarray}
Finally the CGF can be simplified as
\begin{eqnarray}
\ln {\cal Z}&=&t_M \int \frac{dE}{2\pi\hbar} \,\ln \det \Bigl\{ I + G_{0}^r \Gamma_L 
G_{0}^a \Gamma_R \Big[(e^{i\alpha}\! -\! 1)f_L \nonumber \\ 
&&+ ( e^{-i\alpha} \!-\! 1) f_R - (
e^{i\alpha} \!+\! e^{-i\alpha} \!-\!2 ) f_L f_R \Big]\Bigr\}.\qquad .
\end{eqnarray}
where $\alpha=\xi_p+ \xi_e E$ and $f_\alpha$ is the Fermi distribution. Note the difference of the signs in the CGF as compared to the phonons.  If we replace $\alpha$ by $(E-\mu_L) \xi$, the resulting formula is for the counting of the heat 
${\cal Q}_L = {\cal H}_L - \mu_L {\cal N}_L$ transferred, where $\mu_L$ is the chemical potential of the left lead.
The CGF obeys the following fluctuation symmetry \cite{fluct-theorem-1}
\begin{equation}
{\cal Z}(\xi_e,\xi_p)= {\cal Z}\big(-\xi_e + i(\beta_R-\beta_L), -\xi_p -i (-\beta_R \mu_R -\beta_L \mu_L)\big).
\end{equation}



\begin{thebibliography}{99}

\bibitem{Caroli} C. Caroli, R. Combescot, P. Nozieres, D. Saint-James, J. Phys. C: Solid St. Phys. \textbf{4}, 916 (1971).  
\bibitem{Meir} Y. Meir and N. S. Wingreen, Phys. Rev. Lett. \textbf{68}, 2512 (1992).

\bibitem{Rego-Kirczenow-1998} L. G. C. Rego and G. Kirczenow, Phys. Rev. Lett. \textbf{81}, 232 (1998).

\bibitem{Segal} D. Segal, A. Nitzan, and P. H\"anggi, J. Chem. Phys. \textbf{119}, 6840 (2003).

\bibitem{Mingo-Yang-2003} N. Mingo and L. Yang, Phys. Rev. B \textbf{68}, 245406 (2003).

\bibitem{Yamamoto-2006} T. Yamamoto and K. Watanabe, Phys. Rev. Lett. \textbf{96}, 255503 (2006).

\bibitem{Dhar1} A. Dhar and D. Roy, J. Stat. Phys. \textbf{125}, 805 (2006).

\bibitem{Wang-prb06} J.-S. Wang, J. Wang, and N. Zeng, Phys. Rev. B \textbf{74}, 033408 (2006). 

\bibitem{Wang-pre07} J.-S. Wang, N. Zeng, J. Wang, and C. K. Gan, Phys. Rev. E \textbf{75}, 061128 (2007). 

\bibitem{WangJS-europhysJb-2008} J.-S. Wang, J. Wang, and J. T. L\"u, 
Eur. Phys. J. B \textbf{62}, 381 (2008). 

\bibitem{Dhar}A. Dhar, Adv. in Phys., \textbf{57}, 457-537 (2008).

\bibitem{Lepri}S. Lepri, R. Livi, and A. Politi, Phys. Rep. \textbf{377}, 1 (2003).

\bibitem{Lebowitz} F. Bonetto, J. L. Lebowitz, and L. Rey-Bellet, ``Fourier's Law: A Challenge to Theorists,'' {\it {Mathematical Physics 2000}} (Imp. Coll. Press, London, 2000).

\bibitem{eduardos-paper}
E. C. Cuansing and J.-S. Wang, Phys. Rev. B \textbf{81}, 052302 (2010); erratum \textbf{83}, 019902(E) (2011).
\bibitem{eduardos-paper1}
E. C. Cuansing and J.-S. Wang, Phys. Rev. E \textbf{82}, 021116 (2010).

\bibitem{baowen-review} N. Li, J. Ren, L. Wang, G. Zhang, P. H\"anggi, and B. Li, arxiv: 1108:6120.

\bibitem{oconell} A. D. O'Connell, M. Hofheinz, M. Ansmann, R. C. Bialczak, M. Lenander, E. Lucero, M. Neeley, D. Sank, H. Wang, M. Weides, J. Wenner, J. M. Martinis, and A. N. Cleland, Nature, \textbf{464}, 697 (2010).



\bibitem{Levitov} L. S. Levitov and G. B. Lesovik, JETP Lett. \textbf{58}, 230 (1993);

\bibitem{Levitov1} L. S. Levitov, H.-W. Lee, and G. B. Lesovik, J. Math. Phys. 
\textbf{37}, 4845 (1996).

\bibitem{Levitov-PRB} L. S. Levitov and M. Reznikov, Phys. Rev. B \textbf{70}, 115305 (2004).

\bibitem{otherworks} W. Belzig and Y. V. Nazarov, Phys. Rev. Lett. \textbf{87}, 197006 (2001).

\bibitem{otherworks1} Y. V. Nazarov and M. Kindermann, Eur. Phys. J. B \textbf{35}, 413 (2003).

\bibitem{otherworks2} K. Sch\"onhammer, Phys. Rev. B \textbf{75}, 205329 (2007); 
J. Phys.:Condens. Matter \textbf{21}, 495306 (2009).  

\bibitem{Klich} I. Klich, in {\sl Quantum Noise in Mesoscopic Physics}, 
NATO Science Series II,  Vol. 97, edited by Yu. V. Nazarov 
(Kluwer, Dordrecht, 2003).

\bibitem{Pilgram} S. Pilgram, A. N. Jordan, and E. V. Sukhorukov, and 
M. B\"uttiker, Phys. Rev. Lett. \textbf{90}, 206801 (2003).

\bibitem{fluct-theorems} K. Saito and Y. Utsumi, Phys. Rev. B \textbf{78}, 
115429 (2008).

\bibitem{Bagrets} D. A. Bagrets and Y. V. Nazarov, Phys. Rev. B
\textbf{67}, 085316 (2003).

\bibitem{Gogolin} A. O. Gogolin and A. Komnik, Phys. Rev. B \textbf{73}, 195301 (2006). 

\bibitem{Urban} D. F. Urban, R. Avriller, and A. Levy Yeyati,
Phys. Rev. B \textbf{82}, 121414(R) (2010).

\bibitem{Gutman} D. B. Gutman, Y. Gefan, and A. D. Mirlin,
Phys. Rev. Lett. \textbf{105}, 256802 (2010).

\bibitem{Flindt-etal-2009} C. Flindt, C. Fricke, F. Hohls, T. Novotn\'{y}, K. Neto\v{c}n\'{y}, 
T. Brandes, and R. J. Haug, PNAS, \textbf{106}, 10116 (2009). 

\bibitem{Gustavsson-etal-2006} S. Gustavsson, R. Leturcq, B. Simovi\v{c},
R. Schleser, T. Ihn, P. Studerus, K. Ensslin, D. C. Driscoll, and A. C. Gossard,
Phys. Rev. Lett. \textbf{96}, 076605 (2006).

\bibitem{measurement} A. A. Clerk, Florian Marquardt, and J. G. E. Harris, Phys. Rev. Lett \textbf{104}, 213603 (2010).

\bibitem{Saito-Dhar-2007} K. Saito and A. Dhar, Phys. Rev. Lett. \textbf{99}, 180601 (2007); Phys. Rev. E \textbf{83}, 041121 (2011).

\bibitem{ren-jie} J. Ren, P. H\"anggi, and B. Li, Phys. Rev. Lett. \textbf{104}, 170601 (2010). 

\bibitem{AAClerk-2011} A. A. Clerk, Phys. Rev. A \textbf{84}, 043824 (2011).



\bibitem{noneq-fluct} G. Gallavotti and E. G. D. Cohen. Phys. Rev. Lett. \textbf{74}, 2694 (1995); C. Jarzynski, ibid. \textbf{78}, 2690 (1997).

\bibitem{fluct-theorem-1}D. Andrieux,  P. Gaspard, T. Monnai, and S. Tasaki,
New. J. Phys. \textbf{11}, 043014 (2009). 

\bibitem{fcs-bijay}
J.-S. Wang, B. K. Agarwalla, and H. Li, Phys. Rev. B \textbf{84}, 153412, (2011).

\bibitem{Esposito-review-2009} M. Esposito, U. Harbola, and S. Mukamel,
Rev. Mod. Phys. \textbf{81}, 1665 (2009).

\bibitem{Hanggi2} M. Campisi, P. H\"anggi, and P. Talkner, Rev. Mod. Phys. {\bf 83}, 771 (2011).


\bibitem{kundu}
A. Kundu, S. Sabhapandit, and A. Dhar, J. Stat. Mech (2011) P03007.


\bibitem{schwinger-keldysh} J. Schwinger, J. Math. Phys. {\bf 2}, 407
  (1961); L.V. Keldysh, Sov. Phys. JETP {\bf 20}, 1018 (1965).

\bibitem{rammer86} See, for a review, J. Rammer and H. Smith, Rev. Mod.
  Phys. {\bf 58}, 323 (1986).


\bibitem{Hanggi3} M. Campisi, P. Talkner, and P. H\"anggi, Phys. Rev. E  \textbf{83}, 041114 (2011).

\bibitem{Hanggi4} M. Campisi, P. Talkner, and P. H\"anggi, Phys. Rev. Lett. \textbf{105}, 140601 (2010).

\bibitem{neumann} J. von Neumann, \textsl{Mathematical Fundations of Quantum Mechanics}, Princeton Univ. Press, Princeton, (1955). 

\bibitem{Feynman-Vernon-1963} R. P. Feynman and F. L. Vernon, Ann. Phys. \textbf{24}, 118 (1963).

\bibitem{Stockburger} J. T. Stockburger and H. Grabert, 
Phys. Rev. Lett. \textbf{88}, 170407 (2002).

\bibitem{Huag} H. Huag and A.-P. Jauho, \textit{Quantum Kinetics in Transport and Optics of Semiconductors}, 2nd ed. (Springer, New York, 2008). 

\bibitem{Buttiker} Ya. M. Blanter and M. B\"uttiker, Physics Reports, \textbf{336}, 2, (2000).


\bibitem{NR} W.H. Press, S.A. Teukolsky, W.T. Vetterling, and
  B.P. Flannery, {\it Numerical Recipes: The Art of Scientific
  Computing}, 3rd ed. (Cambridge, New York, 2007).


\bibitem{Rubin} R. J. Rubin and W. L. Greer, J. Math. Phys. {\bf 12}, 1686 (1971).

\bibitem{Weiss} U. Weiss, {\it{Quantum Dissipative Systems}}, 2nd edn. (World Scientific, 1999).

\bibitem{Nelson} H. -P. Breuer and F. Peteruccione, {\it{The Theory of Open Quantum Systems}}, Oxford University Press, Oxford, 2002.


\bibitem{ep2} M. Esposito, K. Lindenberg, and C. Van den Broeck, New. J. Phys. \textbf{12}, 013013 (2010). 

\bibitem{ep1} S. Deffner and E. Lutz, Phys. Rev. Lett, \textbf{107}, 140404 (2011).



\bibitem{JE} C. Jarzynski, Phys. Rev. Lett. \textbf{78}, 2690 (1997).
\bibitem{talkner2008} P. Talkner, P. S. Burada, and P. H\"anggi,
Phys. Rev. E \textbf{78}, 011115 (2008).

\bibitem{driven-bijay}
B. K. Agarwalla, J.-S.Wang, and B. Li, Phys. Rev. E \textbf{84}, 041115 (2011).

\bibitem{Zeng} N. Zeng, Ph.D thesis, National Univ. Singapore (2008). \\
At http://staff.science.nus.edu.sg/\char`~phywjs/NEGF/negf.html.
 
\bibitem{Brouwer} P. Brouwer, \textit{Theory of Many-Particle Systems} (Lecture notes for P654, Cornell University, Spring 2005).

\bibitem{Bernard} D. Bernard and B. Doyon, arxiv: 1105.1695.

\bibitem{Hanggi1} F. Zhan, S. Denisovm, and P. H\"anggi, Phys. Rev. B \textbf{84}, 195117 (2011).





\end{thebibliography}
\end{document}